\documentclass[preprint,aps,tightenlines,floatfix,showpacs]{revtex4}
\usepackage{graphics}
\usepackage{bm}
\usepackage{amsmath}
\usepackage{amssymb}

\begin{document}

\title{Low excitation structure of $^{10}$B probed
by scattering of electron and of 197 MeV polarized protons}

\author{K. Amos} 
\email{amos@physics.unimelb.edu.au}
\affiliation{School of Physics, The University of Melbourne, 
Victoria 3010, Australia}

\author{S. Karataglidis}
\email{S.Karataglidis@ru.ac.za}
\affiliation{Department of Physics and Electronics,
Rhodes University, P.O. Box 94, Grahamstown 6140, South Africa}

\author{Y. J. Kim}
\email{yjkim@cheju.ac.kr}
\affiliation{Department of Physics, Cheju National University,
Jeju City, Jeju-Do 690-756, Republic of Korea.}

\date{\today}

\begin{abstract}
Cross-section and analyzing power data from 197 MeV  $(p,p')$ 
scattering and  longitudinal and transverse form factors  for 
electron scattering to low lying states in $^{10}$B have been 
analyzed as tests of the structure of the nuclear states when 
they are described using a no-core   $(0+2)\hbar\omega$ shell 
model.        While the results obtained from the shell model 
clearly show the need of other elements, three-body forces in 
particular, to explain the observed spectrum,  the reasonable 
level of agreement obtained in the analyses of the scattering 
data suggest that the wave functions from our shell model with
only a two-body potential are credible.    Any changes to the 
wave functions with the introduction of three-body  forces  in 
the shell model Hamiltonian  therefore  should  be relatively 
minor.
\end{abstract} 

\pacs{21.10.Hw,25.30.Dh,25.40.Ep,25.80.Ek}
\maketitle

\section{Introduction}
In  a recent  paper~\cite{Be05}, cross  sections and  spin observables
were measured  for the elastic  and inelastic scattering  of polarized
protons from $^{10}$B at an energy of 197 MeV.   In addition data from
the charge exchange $(p,n)$ reaction to the ground state of ${}^{10}$C
were  taken  at  186   MeV~\cite{Wa93};  which  ground  state  can  be
considered as the  isobaric analogue of the $1^+;  T=1$ 1.74 MeV state
in $^{10}$B.  Complementing these  proton   scattering data  are those
from measurements \cite{Ci95} of  the longitudinal and transverse form
factors of  electron scattering from $^{10}$B.    Such a complementary
set  of data provides  an opportunity  to assess  the quality  of model
structures of $^{10}$B, if one has appropriate  means  to analyze that
data. For  incident protons  of energies $\sim  200$~MeV,  elastic and
inelastic  scattering  observables  has  been predicted  well  with  a
$g$-folding  model for the  optical potential  and the  distorted wave
approximation (DWA)  built with the same  effective two-nucleon ($NN$)
interaction~\cite{Am00}.  The electron  scattering  form factors  from
such a light mass nucleus also have been predicted well when allowance
is made of a number of corrections and effective operators~\cite{Ka95}
within  the Born  approximation. Crucial  in finding  good predictions
with  these  reaction  models has  been  the  use  of very  good  wave
functions for the states of  the target nuclei. With light mass nuclei
there are many models that give  such and herein we use one: a no-core
complete  $(0+2)\hbar\omega$ shell model using the fitted interactions 
of Millener and Kurath \cite{Wa89}.

What  makes  $^{10}$B   a  difficult,  and  at  the   same  time  very
interesting, nuclear target in  analyses of proton scattering data, is
that it has a ground state  spin of $3^+$. In all reactions then, save
for the  excitation of $0^+$ states but  including elastic scattering,
multiple angular  momentum transfer values are  possible. With elastic
scattering  in  particular,  all  transfer  values from  0  to  6  are
possible.

 There are a number of  studies~\cite{Am00,Ha03} that show the need to
 use large space models  of structure to adequately analyze scattering
 data.  For light  mass nuclei  in particular,  now that  no-core, and
 complete basis, shell model evaluations are viable, it is of little
 use  to restrict  evaluations  of structure  and/or  scattering to  a
 $0\hbar\omega$  scheme  despite  the   convenience  of  doing  so  in
 calculations. Such has been known  for decades of course, being shown
 in the guise of the large effective charges required with such simple
 models to match electromagnetic transition data. Worse is that larger
 space calculations bring into  the nuclear state descriptions, single
 particle  wave functions  that have  more nodes.  Such  can influence
 properties which are (linear) momentum dependent,   such  as electron  
 scattering  form  factors.   Scaling electron scattering form factors
 to  find the  $B(E2,q)$~\cite{Am00}  demonstrates that  most  clearly.
 Concerning the data of interest,  Betker {\it et al.} \cite{Be05} and
 Cichocki {\it et al.}~\cite{Ci95}   acknowledge the problems of using
 the  limited-space, shell model  wave  functions that  they chose  in
 their analyses. The quite diverse scalings they require to match data
 reflect that. Their results  indicate  the implicit momentum transfer 
 dependence of effective charges also.
 
   Besides the limited structure used, there are other features of the
 previous analyses of the  proton scattering data~\cite{Be05} that are
 of concern. Of those, a major  one is that the distorted wave impulse
 approximation (DWIA) was used. For  incident energies to 200 MeV that
 approximation is  not really appropriate~\cite{Am00}.  First there is
 the  associated   loss  of,  or  gross   approximation  to,  exchange
 scattering  amplitudes. Also  phenomenological,  local optical  model
 potentials  were used to  determine the  distorted wave  functions of
 relative  motion and  those are  known to  be too  large  through the
 nuclear volume due to  inadequate representation of nonlocal effects.
 Often  the  argument  is used  that  a  quality  fit to  the  elastic
 scattering  data  justifies  the  use  of the  relative  motion  wave
 functions generated from phenomenological, local, optical potentials.
 But such fits only require  specification of a suitable set of phase
 shifts and they are determined  from the asymptotic properties of the
 distorted  waves. The  credibility  of the  distorted wave  functions
 through the volume of the nucleus, properties needed in evaluation of
 inelastic scattering amplitudes,  cannot be assured thereby. Finally,
 the  impulse approximation does  not give,  or approximate  well, the
 important  effects due to  specific knock-out  (exchange) amplitudes.
 Even  at  200  MeV,  such  have momentum  transfer  properties  quite
 different  to those of  the direct  scattering matrix  elements, and,
 worse,  often  the   direct  and  exchange  amplitudes  destructively
 interfere~\cite{Am00}. Given the large  set of uncertainties in those
 analyses,  one can  have little  confidence about  conclusions drawn,
 whether about the structure of the target or of a need for additional
 processes such  as  channel  coupling. Thus we have  reanalyzed the
 data~\cite{Be05} using  a $g$-folding model of  the optical potential
 (for  $I=0$ contributions  to  elastic scattering)  and  the DWA  for
 inelastic scattering. The  non-zero angular momentum contributions to
 elastic scattering have been evaluated  also in the DWA. We have used
 the   $g$-folding  model~\cite{Am00}   for  the   (nonlocal)  optical
 potentials  with  the Melbourne  effective  $NN$  force defining  the
 $g$-matrices. Raynal's  DWBA98 code~\cite{Ra98}, which  allows use of
 that  medium, complex,  and  energy dependent  mix  of central,  $NN$
 spin-orbit,   and $NN$  tensor  forces  has been  used  to give  most
 results of proton induced scattering.
 
 We have  not sought  to make a  coupled-channel study. At  much lower
 energies,  where  discrete  state  effects  are  known  to  influence
 scattering, a  coupled-channel model  of scattering is  essential. An
 appropriate one, which ensures  that the Pauli principle is satisfied
 even  with a  collective  model prescription  for  the coupling,  now
 exists and has  been used to explain compound  nucleus structure even
 in exotic, radioactive  light mass systems~\cite{Am03,Ca06}. However,
 for medium energies, such as  at 197~MeV we consider herein, coupling
 between the low excitation energies  in the target is not expected to
 be important, nor has there been  any need for such when a good model
 of structure, and a reasonable  $NN$ force, were used in evaluations.
 That is so at least for cross sections  usually  greater  than  about 
 0.1~mb/sr. There has been a number of papers dealing  with scattering
 using  coupling to  the  continuum, the  so-called  CDCC method.  The
 results have been quite good but  there are a number of problems with
 the approach as it is to date. First, and perhaps most crucially, the
 evaluations  do  not  treat   the  effects  of  the  Pauli  principle
 adequately.   While  there  have   been  attempts   using  equivalent
 localizations of  those effects, the true non-locality  caused by the
 indistinguishableness of the emergent  nucleon with those left in the
 target  gives  scattering  amplitudes  that have  different  momentum
 transfer properties to those of  the direct scattering ones (by which
 the emergent nucleon is that incident on the target). Essentially one
 must use  the full one-body  density matrix elements  (OBDME) of  the
 target and not just the diagonal reduced elements whose sum gives the
 density itself.  The other problem with  the CDCC as  it is presently
 established,  is   that  the  discretization  of   the  continuum  is
 arbitrary, or  at best  linked to very  scanty information  about the
 continuum spectrum  of the target. Some  time ago, it  was shown that
 specific  properties (the  giant resonances)  would  influence proton
 scattering for energies of  protons that coincide with the excitation
 energies of  those resonances in  the target~\cite{Ge75}. Thus  we do
 not dispute a  role of coupled channels in  a scattering process, but
 we are convinced that such  are a requirement when there are specific
 (collective  and not  too spread)  states  in the  target nucleus  at
 excitation in the vicinity of the incident energy value.

  For  electron  scattering  form  factors  we assume  that  the  Born
 approximation,  suitably  adjusted, and  for  $\sim  200$~MeV  proton
 scattering  we  assume  that  the  $g$-folding and  DWA  models,  are
 appropriate to use in data  analyses. In the next section, details of
 the   structure   assumed  for   $^{10}$B   are   given.    Then   in
 Sec.~\ref{Results} we present and discuss the results of our analyses
 of  the scattering  data  while  the conclusions  we  draw are  given
 thereafter in Sec.~\ref{Conclusions}.

\section{Details of the structure assumed for ${}^{10}$B}
\label{Structure}

 Most  studies  needing  the  nucleon based  properties  of  so-called
 $0p$-shell nuclei,  use $0p$-           or at best  $psd$-shell model
 information~\cite{Co65,Mi75}.  Such are  known to  be limited  and to
 give wave functions with which  large effective charges are needed to
 map measured  electromagnetic transition rates. That is  not the case
 now with current larger space, no-core, calculations of structure; as
 has been used for $^{12}$C~\cite{Ka95}. Thus we used the complete $(0
 + 2)\hbar\omega$ space with  the MK3W interaction \cite{Wa89} and the
 OXBASH code~\cite{Br86} to specify the spectrum and wave functions of
 $^{10}$B.

\subsection{The model of structure of $^{10}$B}

\begin{figure}
\scalebox{0.7}{\includegraphics*{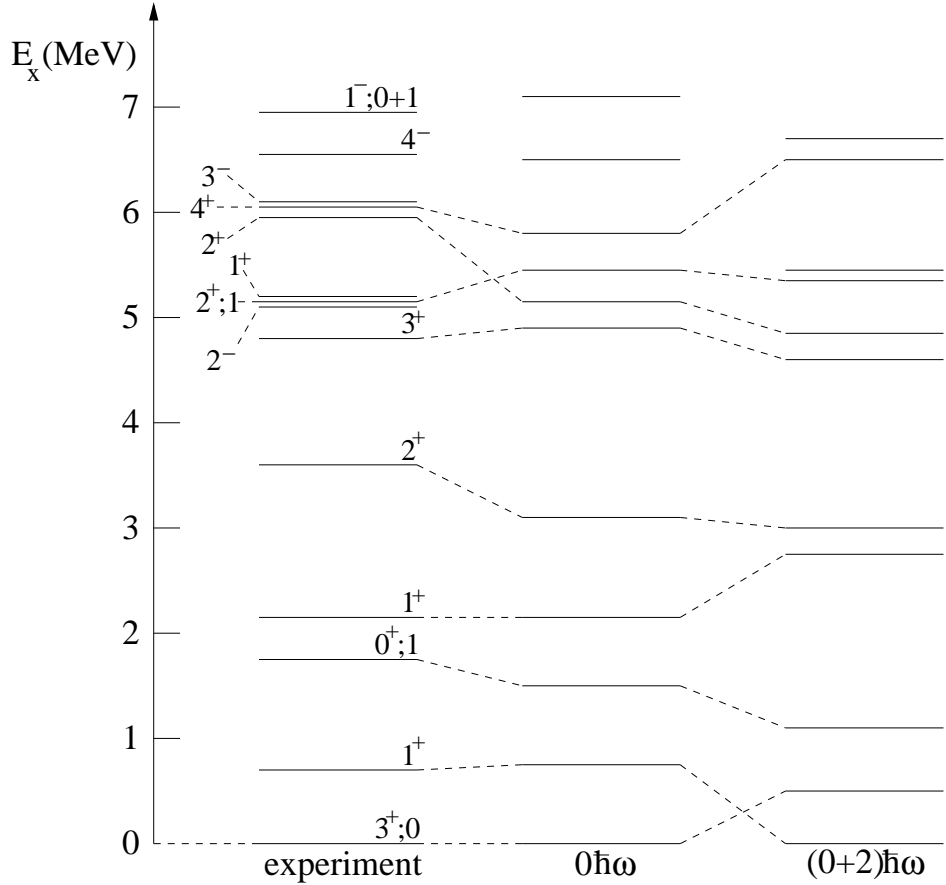}}
\caption{\label{Fig1-amos} The low excitation spectrum  of
  $^{10}$B. The results of the $0\hbar\omega$ and $(0+2)\hbar\omega$
  shell model calculations were made using the CK(8-16)23BME and MK3W interactions,
respectively.}
\end{figure}

 The  spectrum  of  $^{10}$B   that  resulted  from  our  shell  model
 calculations  is  compared  with   the  known  one  \cite{TUNL04}  in
 Fig.~\ref{Fig1-amos}.
While the energies obtained from the shell model calculation  are   in 
good agreement with those observed,     the shell model gives a ground 
state with $J^\pi;T = 1^+;0$ in contrast to  the observed ground state 
of $3^+;0$.                  This is consistent with the result of the 
\textit{ab initio}  shell model calculation of Caurier \textit{et al.} 
\cite{Ca02}, who used a shell model also  with  a  two-body  potential 
only, albeit one obtained directly from the nucleon-nucleon force. The 
inversion may be rectified by the inclusion of a  three-body potential 
in the shell model Hamiltonian~\cite{Ca02}.  This was confirmed by the 
QMC calculations of Pieper, Varga and Wiringa~\cite{Pi02},    but with 
the caveat that the right three-body force had to be used.     We also 
note that in a pure $0\hbar\omega$  shell  model  using  the Cohen and 
Kurath CK(8-16)2BME interaction~\cite{Co65}      we obtain the correct 
$3^+$ ground state, as shown in Fig.~\ref{Fig1-amos}.    The mixing of 
the $2\hbar\omega$ components gives rise to the inversion. However, as
the wave functions from the $0\hbar\omega$ model would then    require 
core polarization corrections to describe the scattering,  we use only 
those wave functions obtained from the $(0+2)\hbar\omega$ model.     A 
correct spectrum may then result also if $4\hbar\omega$ components are
admitted into the shell model space, such as was the case for $^{16}$O 
\cite{Ha90}.

       Our calculation of the $^{10}$B spectrum is very similar to the 
$2\hbar\omega$ calculation performed by       Cichocki \textit{et al.} 
\cite{Ci95}         wherein reasonable agreement was obtained with the 
observed spectrum.   However, there are no spin-parity assignments for 
the low-lying states obtained from their shell model   (Fig. 2 of Ref. 
\cite{Ci95}). Direct comparison with the results of our calculation is 
therefore not possible.

 As a test of the model structure, we calculated the quadrupole moment 
of the ground state      (actually the $3^+_1$ state in our spectrum), 
along with the  $B(E2)$  values  for  several  transitions  among  the 
low-lying states.        We list those in Table~\ref{quads} along with 
comparisons to the $8\hbar\Omega$ results of   Caurier \textit{et al.} 
\cite{Ca02}.
\begin{table}
\begin{ruledtabular}
\caption{\label{quads} 
Quadrupole moment (in $e$fm$^2$) of $^{10}$B and the $B(E2)$ values 
(in $e^2$fm$^4$) for the transitions in $^{10}$B as listed. Comparison 
is made with the $8\hbar\Omega$ results of Caurier \textit{et al.} 
\cite{Ca02}.  The data are from Ref. \cite{Aj88}.}
\begin{tabular}{clll}
Observable & \multicolumn{1}{c}{Expt} 
& \multicolumn{1}{c}{Ref. \cite{Ca02}} &
\multicolumn{1}{c}{Present work} \\
\hline
$Q(3^+_1)$ & $+8.472(56)$ & $+6.799$ & $+6.708$ \\
$B(E2;1^+_1 \rightarrow 3^+_1)$ & $4.13 \pm 0.06$ & 4.512 & 3.185 \\
$B(E2;1^+_2 \rightarrow 3^+_1)$ & $1.71 \pm 0.26$ & 0.163 & 0.270 \\
$B(E2;1^+_2 \rightarrow 1^+_1)$ & $0.83 \pm 0.40$ & 3.742 & 1.172 \\
$B(E2;3^+_2 \rightarrow 1^+_1)$ & $20.5 \pm 2.6$ & 4.754 & 9.057 \\
\end{tabular}
\end{ruledtabular}
\end{table}
   The value of the quadrupole moment from our shell model calculation 
compares favorably with that of Caurier \textit{et al.},      and both 
results are in reasonable agreement with the experimental value.   The 
$B(E2)$  values  for  the listed transitions vary somewhat compared to 
the other model results.     While the $B(E2;1^+_1 \rightarrow 3^+_1)$ 
value from our model is lower compared to both the other result    and 
experiment, our other $B(E2;1^+_2 \rightarrow 3^+_1)$ result  compares 
far more favorably to the  experimental  value,  for  this  relatively 
weak transition.    Our results also agree far more favorably with the 
experimental values. This may indicate the underlying problem with the 
$G$-matrix interaction used by Caurier \textit{et al.}  It restricts 
the long-range correlations to two-body correlations, neglecting terms
of higher order.     The problem stems from the neglect of part of the 
excluded space in the development of the interaction~\cite{Ka08}.

\subsection{Wave functions, one body density matrix elements, and 
transition amplitudes}

Use of the model spectroscopy for $^{12}$C,      in analyses of medium 
energy proton inelastic scattering cross sections and analyzing powers 
permitted an identification of $J^\pi;T$ values for states in $^{12}$C 
that hitherto had uncertain assignments.  As a complete basis was used 
(for the $(0+2)\hbar\omega$ case at least),       there is no spurious 
center of mass motion in the state specifications.  Hence our interest 
in application to the measured data of electron and    197~MeV protons 
scattering from $^{10}$B.

With either probe,  for form factors of electron scattering and  cross 
sections from proton scattering,   we use a nucleon based approach for 
which both single-nucleon bound-state wave functions and  OBDME   from 
the structure model are required.  However, while harmonic oscillators 
were used in the shell model to determine those OBDME,   in scattering 
calculations we chose Woods-Saxon (WS) wave functions   for the single 
nucleon bound-state wave functions.   Their use previously~\cite{Ka95} 
gave better predictions of scattering observables from $^{12}$C   than 
did use of harmonic oscillator wave functions.        The same binding 
energies of states used in the $^{12}$C data analyses   have been used 
for those in  $^{10}$B.

The OBDME arise in formulation of scattering amplitudes.           The 
specification of the electron scattering form factors we evaluate  has 
been published~\cite{Ka95,Fo66} and for electron scattering    between 
nuclear states $J_i$ and $J_f$ involving angular momentum transfer $I$, 
they have the form 
  \begin{equation}
    \left| F^{\xi}_I(q) \right|^2 = \frac{1}{2J_i+1} \left(
    \frac{4\pi}{Z^2} \right) \left| \left\langle J_f \left\|
    T^{\xi}_I(q) \right\| J_i \right\rangle \right|^2,
  \end{equation}
where $\xi$ selects the type, i.e. longitudinal, transverse  electric, 
or transverse magnetic.      Assuming  one-body operators, the reduced 
matrix elements may be expressed in the form,
  \begin{equation}
    \left\langle J_f \left\| T^{\xi}_I(q) \right\| J_i \right\rangle
    = \frac{1}{\sqrt{2I+1}}\text{Tr}(SM),
  \end{equation}
where  $S$   is  the   matrix  of  one-body   transition  densities,
$S_{j_1 j_2 I}$, defined as
  \begin{equation}
   S_{j_1 j_2 I} = \left\langle J_f \left\| \left[ a^{\dag}_{j_2}
      \times \tilde{a}_{j_1} \right]^I \right\| J_i \right\rangle.
  \end{equation}
$M$  denotes the  matrix elements  of the  one-body  longitudinal   or
transverse electromagnetic operators  for each allowed   particle-hole
excitation ($j_1-j_2^{-1}$). Bare  operators  are used for the results 
presented herein,    and explicit meson-exchange-current (MEC) effects 
are ignored.   However,  MEC have been incorporated implicitly  in the  
transverse electric  form  factors  in the long-wave  limit by   using 
Siegert's  theorem \cite{Fr84}.     That serves  to introduce into the 
transverse  electric form factor an explicit dependence  on the charge 
density,  through the use of the  continuity equation. Also the Darwin 
correction  has  been  included  in  the   Coulomb  operator  for  the 
longitudinal form factor.

To  predict  the differential  cross  sections  for  both elastic  and
inelastic scattering  from the Carbon isotopes we use the  microscopic
$g$-folding  model  of the  Melbourne  group~\cite{Am00}.  That  model
begins with  the $NN$  $g$-matrices for  the  interaction of a nucleon
with infinite nuclear matter.        Starting with the BonnB free $NN$ 
interaction~\cite{Ma87},   those  $g$-matrices  are  solutions  of the 
Brueckner-Bethe-Goldstone equations  for  infinite  nuclear matter  of
diverse densities ($\propto k_F^3$, when $k_F$ is the Fermi momentum).
Both Pauli blocking of states and an average background mean field  in 
which the nucleons move are involved and lead to $g$-matrices that are  
complex, energy and medium (density) dependent. They are also nonlocal 
in that the solutions for different partial waves  reflect a tensorial
character. Such can be, and have been, used directly in momentum space 
evaluations of $NA$ (elastic) scattering~\cite{Ar96}, but we prefer to 
analyze data using a coordinate space representation. For this, and to
make use of the program   suite   DWBA98~\cite{Ra98}, the $g$-matrices 
must be mapped,  via a double  Bessel transform to equivalent forms in
coordinate  space.             Folding  those effective  interactions, 
$g_{\text{eff}}$, with the density-matrices of the target then  yields
a complex, nonlocal, density-dependent, nucleon-nucleus ($NA$) optical  
potential  from which the elastic scattering observables are obtained.
Full  details  of  this  prescription  can  be  found  in  the  review 
article~\cite{Am00}.

Inelastic nucleon scattering,    and non-zero multipole amplitudes  of 
elastic scattering, are calculated within the DWA using  the effective
coordinate space $g$-matrices  ($g_{\text{eff}}$)  as  the  transition 
operator.  Again all details are given in the review~\cite{Am00}.  The
transition amplitude has the form
\begin{equation}
T^{M_f M_i \nu' \nu}_{J_fJ_i}(\theta) = 
  \left\langle
  \chi^{(-)}_{\nu'} \right| \left\langle \Psi_{J_fM_f} \right| A
  g_{\text{eff}}(0,1) \mathcal{A}_{01} \left\{ \left|
  \chi^{(+)}_{\nu} \right\rangle \left| \Psi_{J_i M_i} \right\rangle
  \right\},
\end{equation}
where $\chi^{(\pm)}$ are distorted wave functions for an  incident and 
emergent nucleon respectively. Those wave functions are generated from 
$g$-folding optical potentials.   Coordinates 0 and 1 are those of the 
projectile and of a chosen struck bound-state nucleon,   respectively,
and ${\cal A}_{01}$  is  a  two-nucleon  antisymmetrization  operator. 
Then, by using a co-factor expansion of the target wave function,  one 
obtains
\begin{multline}
T^{M_f M_i \nu' \nu}_{J_fJ_i}(\theta) = 
    \sum_{\alpha_1 \alpha_2
    m_1 m_2} \sum_{JM} \frac{(-1)^{j_1 - m_1}}{\sqrt{2J_f + 1}}
\\
 \left
    \langle 
  \left. j_2 \, m_2 \, j_1 \, -m_1 \right| J_f \, M_f \right\rangle
  \left\langle \left. J_i \, M_i \, J \, M\, \right| J_f \, M_f
  \right\rangle
\\   
\left\langle J_f \left\| \left[ a^{\dag}_{j_2} \times
    \tilde{a}_{j_1} \right]^J \right\| J_i \right\rangle
\;  \left\langle \chi^{(-)}_{\nu'}(0) \right| \left\langle
  \varphi_{\alpha_2}(1) \right|
\\
\times 
 A g_{\text{eff}}(0,1)
\mathcal{A}_{01}
  \left\{ \left| \chi^{(+)}_{\nu}(0) \right\rangle \left|
  \varphi_{\alpha_1}(1) \right\rangle \right\}
\end{multline}
for an angular momentum transfer  $J$, and $\alpha$ denotes the set of
single-particle quantum numbers $\{ n, l, j, m_{\tau}\}$, where $\tau$
is the nucleon isospin.   Thus  the scattering amplitudes are weighted 
sums of two-nucleon amplitudes with those weights being the transition 
OBDME, $S_{j_1 j_2 I}^{J_i J_f}$.      With the $g$-folding potentials 
defining the distorted waves,   and the $g_{\text{eff}}$ also taken to 
be the transition operator,  the problem reduces to one  of specifying 
the structure of the target.

\subsection{Observables}

Besides differential cross sections, spin observables of diverse kinds
may be measured when one has polarized beams and the means to   detect 
the polarization of particles.         Differential cross sections and 
analyzing powers for nucleon-nucleus scattering are easily defined  in 
terms of the above amplitudes by
\begin{align}
\frac{d\sigma}{d\Omega} &=
\frac{1}{2(2J_i+1)} \sum_{M_f M_i \nu' \nu}
\left| T_{J_f J_i}^{M_f M_i \nu' \nu}(\theta) \right|^2\ ,
\nonumber\\
A_y &=
Tr\left[ {\mathbf T}^\dagger_{J_f J_i}(\theta)
\ {\mathbf \sigma}_y\ 
{\mathbf T}_{J_f J_i}(\theta) \right]
\ /\ \frac{d\sigma}{d\Omega}
\ ,
\end{align}
where the $y$-axis is directed normal to the scattering plane.

 For most other spin observables, it is more convenient to specify the 
scattering amplitudes in a helicity formalism~\cite{Ja59,Ra72}      in 
which the spin of the particle is projected onto its momentum and  the 
angular momentum of the target is projected onto the reverse direction.  
For the target,    the definition of the helicity as projection of the 
spin  on  the  impulsion  of  the  target  implies a quantization axis 
opposite to that of the particle,       at least in the center of mass 
system.  All details are given in the review~\cite{Am00},  and so only 
a brief summary follows.

  With the axes of quantization along $\mathbf{k}_i(\mathbf{k}_f)$ for 
the  initial  (final)  scattering particle states, helicity amplitudes 
relate to those specified above by simply  the  action  of  a  reduced 
rotation matrix element.  With $J$ being the angular momentum transfer 
quantum number, these helicity amplitudes are
\begin{equation}
T^{hel}_{M_f M_i \nu' \nu}(\theta) =
\sum_J T_{J_i J_f}^{M_f M_i \nu' \nu}(\theta) \ 
r^{(J)}_{M_f-\nu',M _i-\nu}(\theta)\ .
\label{JN30}
\end{equation}
       The utility of the helicity formulation is that all observables
defined with respect to the outgoing  center  of  mass momentum can be
defined without further rotations. Without limiting spin values,   all
observables can be described~\cite{Am00}  with simple tensor operators
$\tau_{\lambda ,\mu }$  whose  matrix  elements in the spin space of a
particle of spin $s (=\frac{1}{2})$ are
\begin{equation}
\left\langle s\,q \left| \tau^s_{\lambda, \mu} \right| s\, q'
\right\rangle = (-1)^{s-q'}\sqrt {2s+1} 
\left\langle s \; q \; s \; -q' \left|  \lambda \; \mu \right\rangle 
\right. \; .
\end{equation}
Similar tensors are given for the target space.     These irreducible,
hermitian,   tensor   operators   are   orthonormal   and      satisfy
$\tau_{\lambda ,\mu } =  (-1)^{\mu }\tau^{\dagger }_{\lambda ,-\mu }$.
Further, all observables can be defined from the coefficients
\begin{equation}
A^{\lambda_3\mu_3\lambda_4\mu_4}_{\lambda_1\mu_1\lambda_2\mu_2} =
\mathrm{Tr}\left\{ T^{hel}(\theta) \left[ 
\tau^{\frac{1}{2}}_{\lambda_1 \mu_1} \otimes 
\tau^{J_i}_{\lambda_2 \mu_2} \right]\right.
\left.
T^{hel}(\theta)^{\dagger} \left[ \tau^{\frac{1}{2}}_{\lambda_3 \mu_3}
\otimes \tau^{J_f}_{\lambda_4 \mu_4} \right] \right\}
\end{equation}
when  one  takes  into  account that the projection of the spin of the
target on the direction of the beam is opposite to its helicity.

In terms of these amplitudes, the differential cross section is defined 
by
\begin{equation}
\frac{d\sigma}{d\Omega} = \frac{1}{2(2J_i+1)} A^{0000}_{0000} \; ,
\end{equation}
and spin observables, generically expressed by $\Gamma(\theta)$,   are
found from 
\begin{equation}
\Gamma(\theta ) A^{0000}_{0000}
= \sum_{\lambda_i \mu_i}
x^{\lambda_3\mu_3\lambda_4\mu_4}_{\lambda_1\mu_1\lambda_2\mu_2}\ 
A^{\lambda_3\mu_3\lambda_4\mu_4}_{\lambda_1\mu_1\lambda_2\mu_2}\ ; 
\end{equation}
the weight coefficients
    $x^{\lambda_3\mu_3\lambda_4\mu_4}_{\lambda_1\mu_1\lambda_2\mu_2}$
specifying each observable.     Such allow consideration of polarized
projectiles, ejectiles, as well as of initial and final targets.    A 
complete set of those, and of the weights defining them, are given in 
the review~\cite{Am00}  and  that  entire  set of observables for the 
scattering of nucleons from nuclei can be evaluated using the  DWBB97 
code~\cite{Ra98}. 

Of interest in this study,  besides the analyzing power $A_y(\theta)$ 
defined above,                  are polarization transfer amplitudes, 
$D_{NN'}(\theta), D_{LL'}(\theta)$, and $D_{LS'}(\theta)$.        The 
subscripts identify axes specified by the momentum vectors of     the 
incident and emergent  nucleons,   $\mathbf{k}$   and   $\mathbf{k}'$
respectively, with
\begin{align}
\hat{\mathbf{N}} &= \frac{\mathbf{k} \times \mathbf{k}'}
{\mathbf{k} \times \mathbf{k}'}\ ,
\nonumber\\
\hat{\mathbf{L}} &= \frac{\mathbf{k}}{k}\ ,
\nonumber\\
\hat{\mathbf{S}} &= {\mathbf L} \times {\mathbf N}\ .
\end{align}
Primed labels refer to the outgoing properties.

Liu {\it et al.}~\cite{Li96}  noted that for the elastic scattering of 
spin $\frac{1}{2}$ particles from zero-spin nuclei,    that there were 
linkages between the five possible  polarization  transfer observables
so that only three would be independent, and that the polarization and 
analyzing powers would equate ($A_y(\theta) = P(\theta)$). Thus
\begin{align}
D_{NN'}(\theta) &= -\frac{1}{A_{0000}^{0000}}
\left[ A_{1100}^{1100} + A_{1100}^{1-100}\right]
\equiv 1 \nonumber\\
D_{LL'}(\theta) &= \frac{1}{A_{0000}^{0000}} A_{1000}^{1000}
\equiv D_{SS'}(\theta) \nonumber\\
D_{LS'}(\theta) &= 
-\frac{\sqrt{2}}{A_{0000}^{0000}}\, A_{1000}^{1100}
\equiv - D_{SL'}(\theta) .
\label{Symcon}
\end{align}
They  noticed  that  these  relationships  also  worked  well for  the
natural parity inelastic transitions they studied.

\section{Discussion of results}
\label{Results}

\subsection{Electron scattering form factors}

Cichocki {\it et al.}~\cite{Ci95} measured longitudinal and transverse
form factors for electron scattering from ${}^{10}$B.    They used six 
energies ranging from 48 to 453 MeV to ascertain those form factors in
the momentum transfer range 0.48 to 2.58 fm$^{-1}$.    Their data have 
been complemented by the (collected) set given for the elastic and the
excitation of the $0^+; T=1$ 1.74 MeV state reported earlier  by Hicks 
{\it et al.}~\cite{Hi88}.   Cichocki {\it et al.}~\cite{Ci95} analyzed 
their form factor data using shell model wave functions.   Their basic 
shell  model  for  positive  parity  states  was  a  $0\hbar\omega$ (a 
$0p$-shell)  model  but  they  also allowed $2\hbar\omega$ corrections  
seeking to explain the lack of strength  (of $C2$ type in  particular) 
that resulted.    Betker \textit{et al.}~\cite{Be05} in their analyses 
of both electron and proton scattering data,  restricted consideration 
of the structure of $^{10}$B to solely the  $0\hbar\omega$ shell model 
and used  oscillator  functions  for  the single-nucleon bound states. 
However,  they  adopted  a  scheme  of  varying the oscillator lengths 
according to reaction and multipolarity, as well as adjusting the size 
of amplitudes  to  find  a  best  fit  to  the  form factors and cross 
sections.     This is a very dangerous scheme to adopt and can lead to 
underlying assumed nuclear Hamiltonians that are quite wrong. Such was 
demonstrated~\cite{Am89}  in regard to the wave functions of $^{14}$N. 
As noted then,          and demonstrated now by the analyses of Betker 
\textit{et al.}~\cite{Be05},    the problem is that the limitations of 
the  $0p$-shell  basis  vary  with  the  state  of  the  system  being 
considered.

 Our use of the complete $(0+2)\hbar\omega$ shell model wave functions
in the analyses does not give rise to such problems.   In addition, we 
do  not  include  any  additional  core-polarization       corrections 
\textit{a posteriori},      preferring instead to use bare charges and 
identify any improvements to the underlying wave functions from better 
model input.

\subsubsection{Ground state form factors}

\begin{figure}[h]
\scalebox{0.7}{\includegraphics*{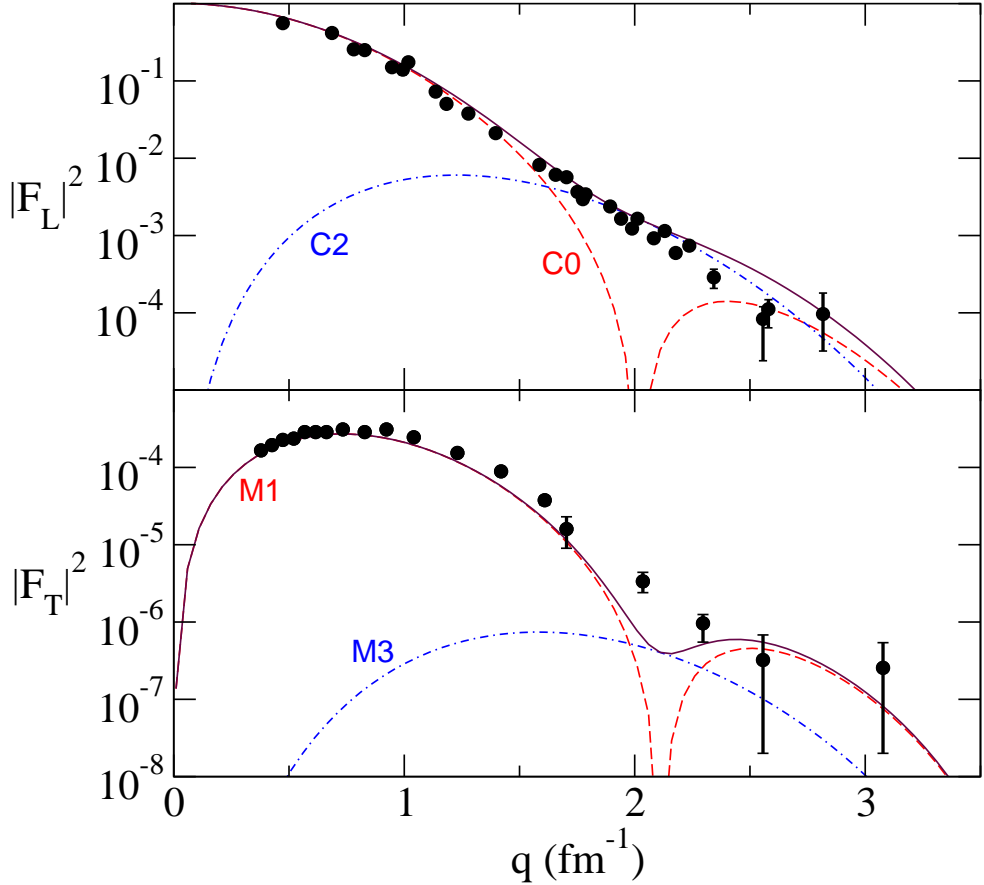}}
\caption{\label{Elect-gs-ff} (Color online)
The longitudinal and transverse form factors for scattering from 
the ground state of $^{10}$B. Data~\cite{Hi88,Ci95} are displayed
by the filled circles.}
\end{figure}

In Fig.~\ref{Elect-gs-ff}, we display the longitudinal $(|F_L|^2)$ and 
transverse $(|F_T|^2)$  form  factors  from  the elastic scattering of 
electrons from $^{10}$B.   The data~\cite{Hi88,Ci95} are compared with
our calculated results (solid curves)   with the dominating components 
of those results as indicated.    In the longitudinal form factor, the
$C0$ and $C2$ terms are shown by the dashed curves      while the most
important elements in the transverse form factor      are the magnetic
dipole ($M1$) and octupole ($M3$) components.   Clearly each component
influences the results at different momentum transfer values.      The 
match to data is very good.
These  results  for  the longitudinal form factor concur in shape with 
those obtained~\cite{Be05}.  However, we require no enhancement of the 
$C2$ term, as required by Betker {\it et al.}~\cite{Be05},  to achieve 
the good agreement with data.   More noticeable though are differences
we find in the transverse form factor results. In this  we concur with 
the assessment made by Hicks {\it et al.}~\cite{Hi88}.           These 
differences we attribute  largely to  changes wrought by considering a 
$(0+2)\hbar\omega$   model   of   the   structure   instead   of   the 
$0\hbar\omega$ model that was used by Betker {\it et al.}~\cite{Be05}. 

\subsubsection{Form factors for excitation of the $1^+$ and $3^+$ 
states}

\begin{figure}[h]
\begin{center}
\scalebox{0.6}{\includegraphics*{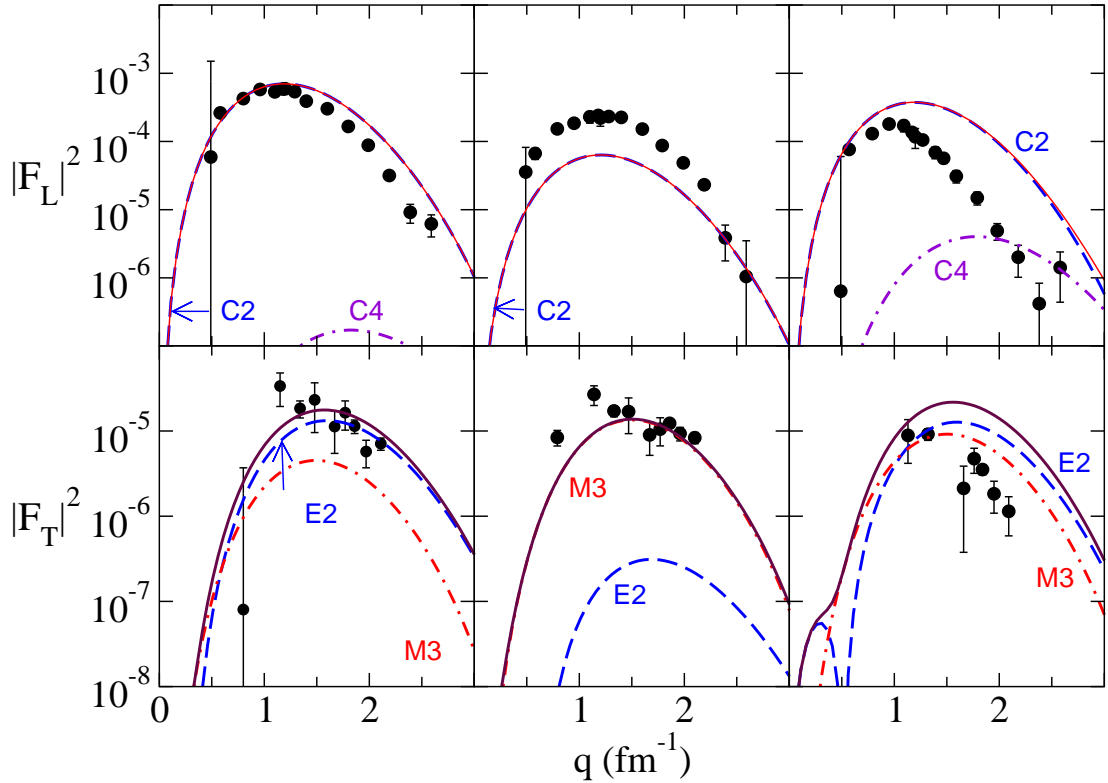}}
\end{center}
\caption{\label{Fig-Elect-1+3+} (Color online)
The longitudinal (top) and transverse (bottom) form factors for 
electron scattering to the $1^+$ 0.718 MeV (left), the $1^+$ 2.154 MeV 
(middle), and the $3^+$ 4.774 MeV (right) states of $^{10}$B.  
Data~\cite{Hi88,Ci95} are displayed by the filled circles.}
\end{figure}

In Fig.~\ref{Fig-Elect-1+3+},     the longitudinal  and transverse form 
factors from electron scattering to the $1^+$  and $3^+$ states     in 
$^{10}$B are shown. The data~\cite{Hi88,Ci95}  are displayed by filled 
circles with  the   longitudinal/transverse  form factors shown in the 
top/bottom segments in this diagram.          The data and results for 
excitation of the $1^+$ (0.718~MeV), the $1^+$ (2.154~MeV),    and the 
$3^+$ (4.774~MeV) states are presented in the  left, middle, and right 
panels respectively.
The  solid  curves  are  the total results being sums over the allowed
multipole contributions.         The longitudinal form factors all are 
dominated by the $C2$ components while $E2$ and $M3$ contributions are 
significant in the transverse form factor evaluations.    The $C2$ and 
$E2$ contributions are displayed by the  long dashed  curves while the 
$M3$ values are shown by the dot-dashed lines.   Cichocki {\it et al.}
\cite{Ci95}  found  similar  results  with  their  analysis  of    the 
longitudinal form factors though their  $2\hbar\omega$  model  results 
lay just under the data for both $1^+$ excitations.    Our results for 
the longitudinal form factor for the $3^+$ 4.774~MeV state agree  with 
that found previously~\cite{Ci95}  but  both calculations overestimate 
observation.

Our results for the transverse form factors are in good agreement with
the data and are the result of mixtures of   $E2$ and  $M3$  multipole 
contributions predominantly.   The $M1$ contributions to the 4.774~MeV 
results exist but are quite small,   effecting small changes in values  
at low-$q$ values ($q \preceq 0.5$~fm$^{-1}$).   We do not display it or 
its effect.     Of note is that the $E2$ and $M3$ contributions are of 
similar size in the transverse form factors for the    $1^+$ 0.718 MeV 
and $3^+$ 4.774 MeV cases,    but the $M3$ is predominant in that form 
factor for scattering to the $1^+$ 2.154 MeV state.

\begin{figure}[h]
\begin{center}
\scalebox{0.7}{\includegraphics*{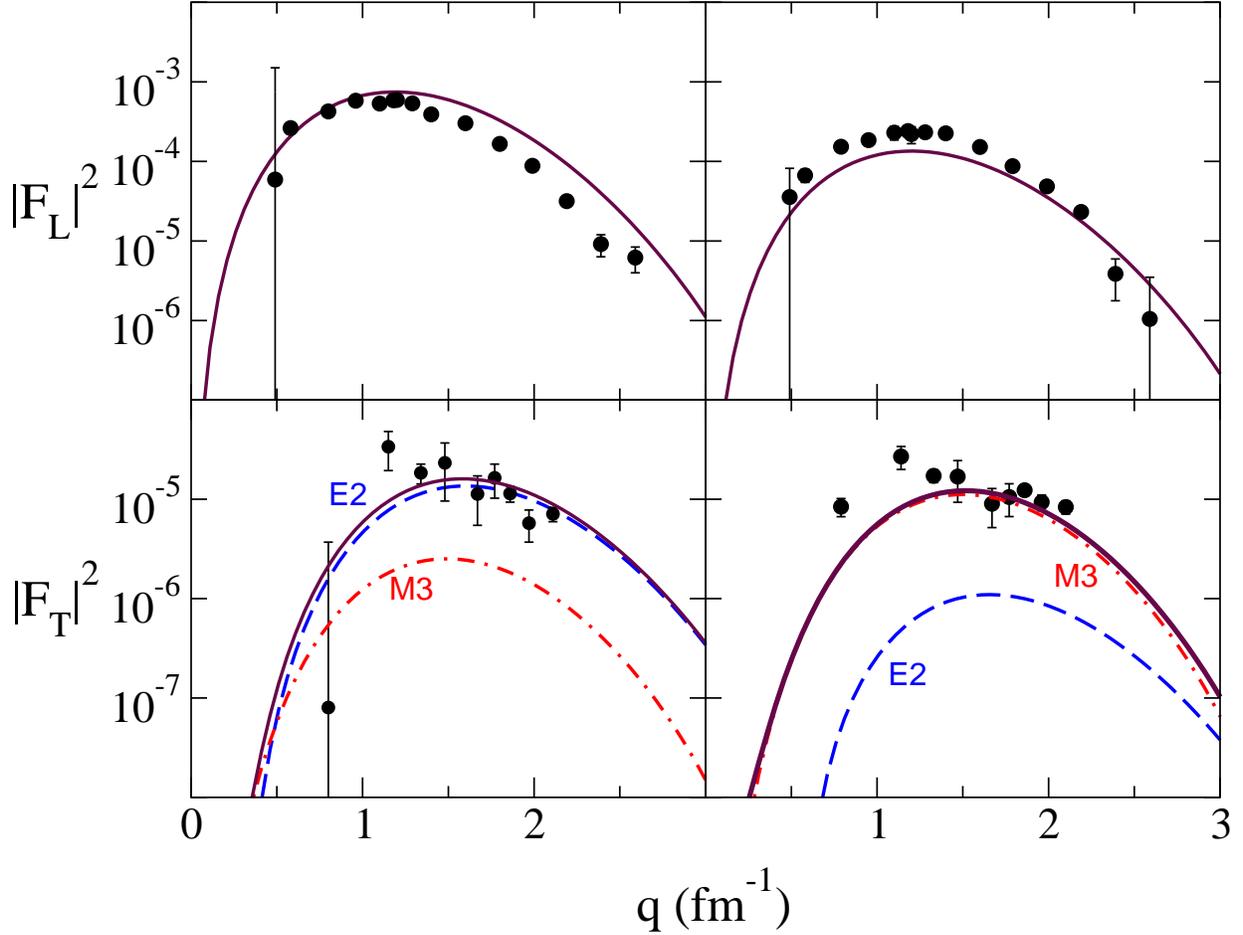}}
\end{center}
\caption{\label{Fig-Elect-mix-1+} (Color online)
The longitudinal (top) and transverse (bottom) form factors for 
electron scattering to the $1^+$ 0.718~MeV (left) and to the the $1^+$ 
2.154~MeV (right) states in $^{10}$B. Results are those from the mixed 
shell model states.}
\end{figure}

As our calculated form factors for the $1^+$ 0.718~MeV state is larger 
than observation  while that for the $1^+$ 2.154 MeV state is smaller,
we considered a description for the two $1^+ (T=0)$ states as a mix of 
those  defined  by  our shell model (designated as $1^+_1$ and $1^+_2$ 
next) to be
\begin{align}
\left| 1^+; 0.718 \right\rangle &= C_1 \left| 1^+_1 \right\rangle
+ \sqrt{1 - C_1^2}\; \left| 1^+_2 \right\rangle
\nonumber\\
\left| 1^+; 2.154 \right\rangle
&= 
\sqrt{1 - C_1^2}\; \left| 1^+_1 \right\rangle
- C_1 \left| 1^+_2 \right\rangle\ .
\end{align}
Varying  the  coefficients  to  define new sets of  OBDME  for the two 
longitudinal  form factor  evaluations  then gave the results shown in 
the top panels of Fig.~\ref{Fig-Elect-mix-1+}.
The coefficient used was $C_1= 0.99$ and the change in the form factor
for the  2.154~MeV state seems dramatic but it must be remembered that 
the axes are linear-logarithmic. With that coefficient, the transverse 
form factors were then evaluated and the results are compared with the
data in the bottom segments of the figure.    The good agreement found 
previously has been retained with the (small)  admixture  states  with 
only the balance between $E2$ and $M3$ contributions being changed.

\begin{figure}[h]
\begin{center}
\scalebox{0.7}{\includegraphics*{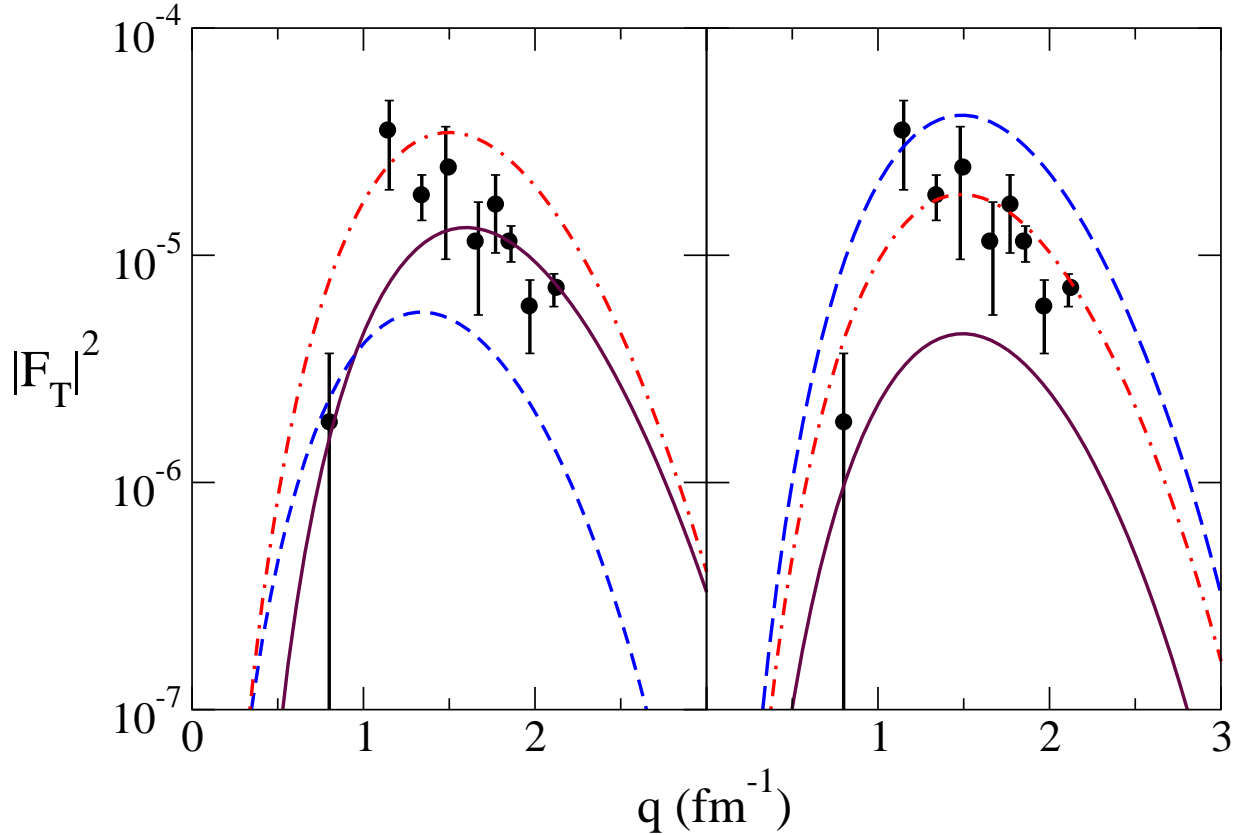}}
\end{center}
\caption{\label{Fig-Elect-1+-pns} (Color online)
The transverse form factors for electron scattering to the $1^+$ 0.718 
MeV state of $^{10}$B showing the separate contributions from the 
protons (dashed line) and neutrons (dot-dashed line) in the target to 
the $E2$ (left) and $M3$ (right) multipolarities.
The total transverse form factor for each multipolarity is shown by
solid curves with the data again displayed by the filled circles.}
\end{figure}

The  comparisons  of results with transverse form factor data in these 
cases (isoscalar, even parity transitions)  is  the  more   remarkable 
since  they  are   results   of   destructive   interference   between 
contributions involving  the protons and the neutrons separately. Such 
is shown in Fig.~\ref{Fig-Elect-1+-pns}.
Both  the  $E2$  (left)  and $M3$ (right) proton and neutron component 
amplitudes destructively interfere to produce the final $E2$ and  $M3$ 
form factors that are displayed by the solid curves.   Thus relatively 
small  variation  in  the  structure  description can have significant 
effects on these results.     Nonetheless the final total result found 
using our $(0+2)\hbar\omega$ model prescription  (the solid curve left 
lower diagram in Fig.~\ref{Fig-Elect-1+3+}),   and in the left diagram 
of Fig.~\ref{Fig-Elect-mix-1+},        is in quite good agreement with 
observation.

\begin{figure}[h]
\begin{center}
\scalebox{0.7}{\includegraphics*{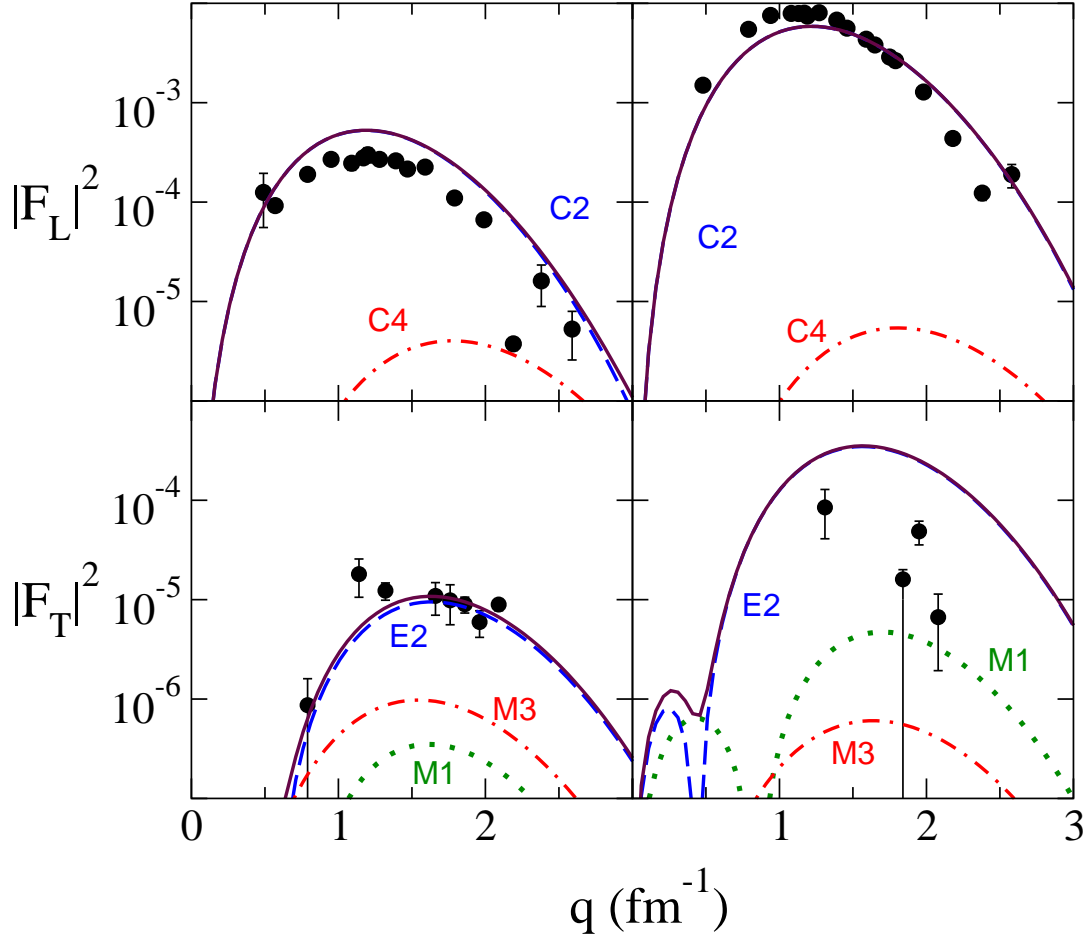}}
\end{center}
\caption{\label{Fig-Elect-2+4+} (Color online)
The longitudinal (top) and transverse (bottom) form factors for 
electron scattering to the $2^+$ 3.587 MeV (left) and 
the $4^+$ 6.025 MeV (right) states of ${}^{10}$B.}
\end{figure}

In Fig.~\ref{Fig-Elect-2+4+} the form factors from electron  scattering 
to the  $2^+$ 3.587~MeV  and  $4^+$ 6.025~MeV  states in  $^{10}$B are 
displayed.   The $2^+$ 3.587~MeV results are shown on the left and the 
longitudinal form factors are displayed in the top segments. Again the 
data~\cite{Hi88,Ci95} are displayed by the filled circles.
\begin{figure}[h]
\begin{center}
\scalebox{0.7}{\includegraphics*{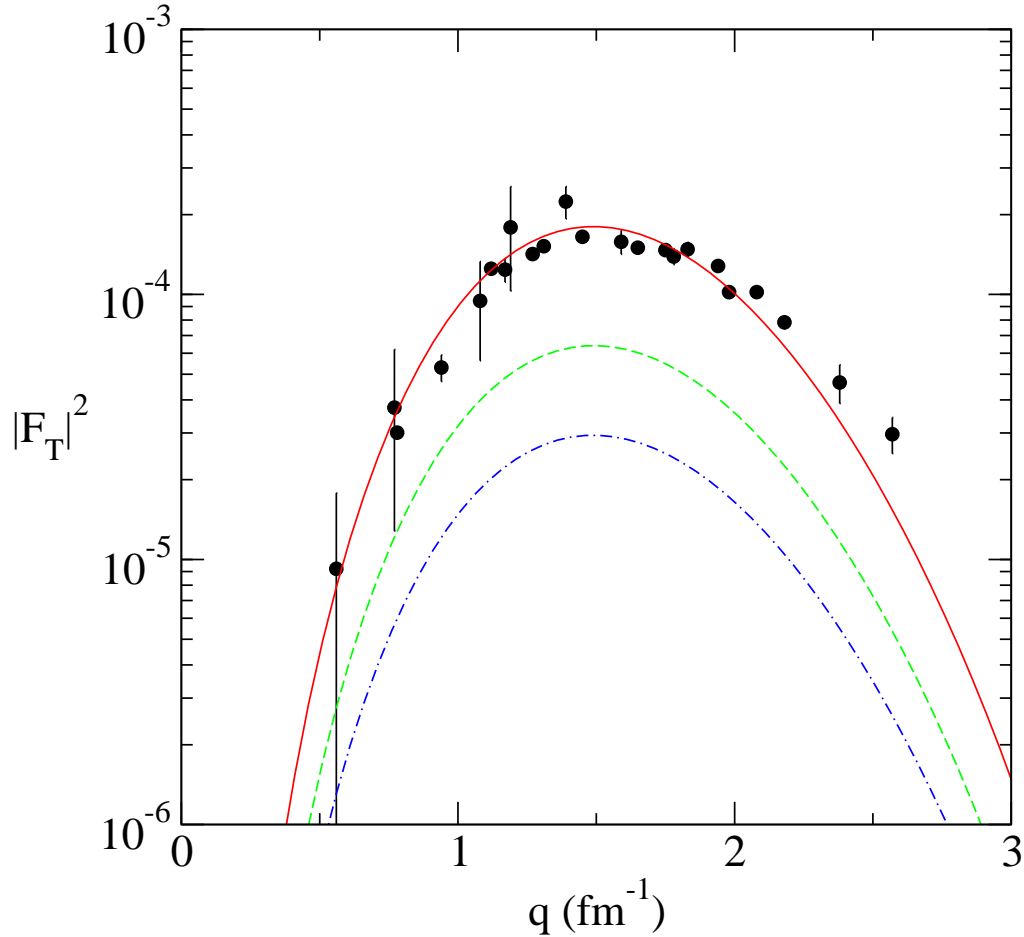}}
\end{center}
\caption{\label{Fig-Elect-0+1.74-tran} (Color online)
The transverse form factor for the $0^+; T=1$ 1.74 MeV state.}
\end{figure}
The  longitudinal  form  factors  are  dominated by the $C2$ multipole
transition  and  our  model  structure  gives  results  in  quite good 
agreement with both data sets.    The small effect of the $C4$ term is 
shown in these plots.  Our result for the $2^+$ transition agrees well 
with that found by Cichocki {\it et al.}~\cite{Ci95} but they required 
sizable core polarization additions to their model to fit  the   $4^+$ 
transition (longitudinal) form factor.  None are needed with our model 
structure. 

The transverse form factors for these   two states shown in the bottom 
segments of Fig.~\ref{Fig-Elect-2+4+},         along with the separate 
multipole contributions.  The results for the form factor of the $2^+$
3.587 MeV  state agrees  quite  well  with the data while that for the
$4^+$ 6.025 MeV  overestimates  the  rather sparse  data so far taken.

Finally we consider the form factor (purely transverse)  measured with 
electron scattering to the $0^+, T=1$ 1.74 MeV state in  the  $^{10}$B 
spectrum.    This is the state we will consider as the analogue to the 
ground state of  $^{10}$C which has been studied~\cite{Wa93} using the 
$(p,n)$ reaction on $^{10}$B.   Form factors for this excitation are 
compared with data in Fig.~\ref{Fig-Elect-0+1.74-tran}.    Therein the 
separate proton and neutron form factors are shown by the  long dashed 
and dashed curves respectively, while their sum, constructive for this
isovector transition, is depicted by the solid curve. The total result 
is in very good agreement with data (filled circles).     This is also
consistent with the  observation~\cite{Ci95}     that very little core 
polarization  correction  is  needed  to describe  this isovector $M3$ 
transition. With our model of structure none is required.

\subsection{Elastic scattering of 197 MeV protons}
\begin{figure}[h]
\begin{center}
\scalebox{0.7}{\includegraphics*{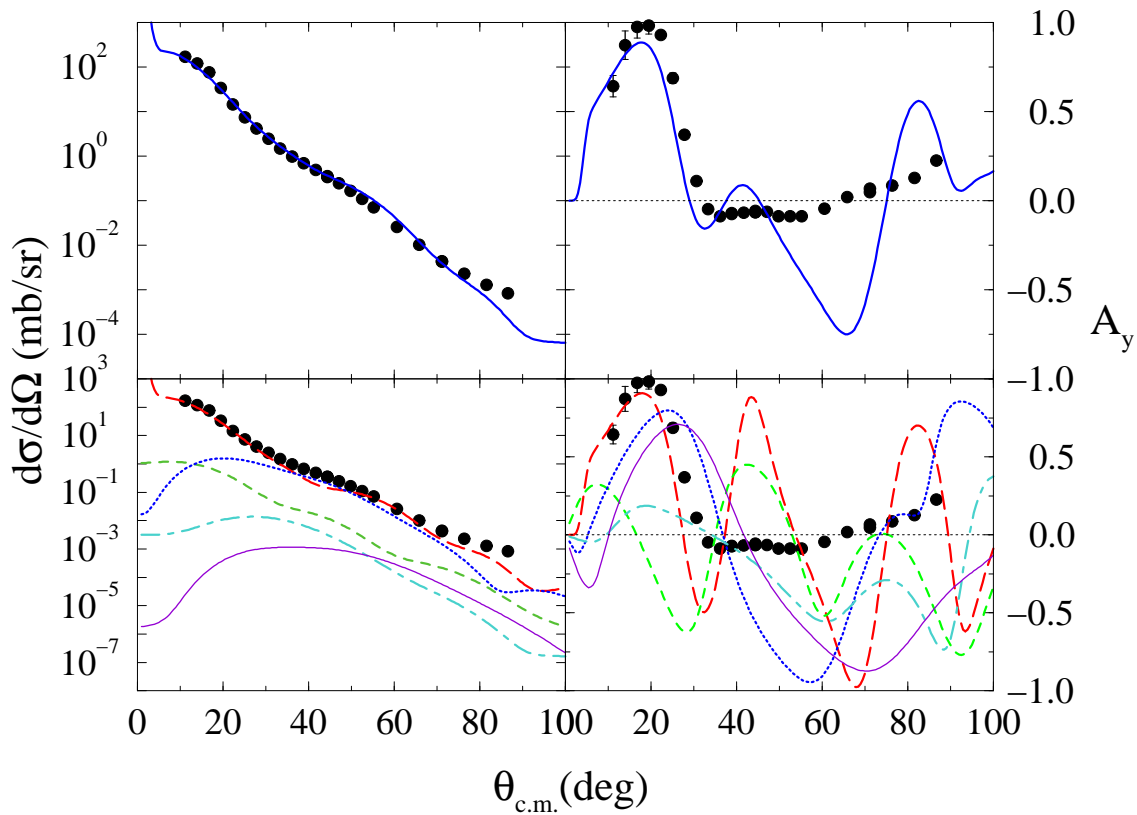}}
\end{center}
\caption{\label{Fig-el-197} (Color online)
The elastic scattering cross section (left) and analyzing
power (right) from 197 MeV polarized proton elastic scattering from 
$^{10}$B. Complete results are compared with data (filled circles)
in the top panels and component contributions are shown in the bottom 
panels.  Details are given in the text.}
\end{figure}

In Fig.~\ref{Fig-el-197} the evaluations of the cross section
and analyzing power for elastic scattering of 197~MeV polarized
protons are compared with the data~\cite{Be05}; the latter being
displayed by the filled circles. 
In the top panels,    the solid lines are the total calculated results
which include contributions from all allowed multipoles.           The  
individual multipole contributions are portrayed in the bottom  panels
by the long dashed lines for $I=0$, the dashed lines for $I=1$,    the 
dotted curves for $I=2$, the dot-dashed curves for $I=3$,      and the 
solid curves for $I=4$.    Note each component analyzing power is that 
relative to the individual component cross section  so their influence 
on the total analyzing power result is only proportionate to       the 
multipole contribution to the cross section strength. As the effect of 
the $I=4$ amplitudes in the total cross section is very minor    while 
the $I=0$ contribution is dominant for the forward angles,  the latter 
is the dominating term in the total analyzing power and the former can 
be neglected.      The other two allowed components, those for angular 
momentum transfer values  $I=5$  and  $I=6$ are much weaker and so are 
not displayed.  Clearly the $I=2$ contributions become significant for 
scattering angles greater than 30$^\circ$ so that the  resultant cross 
section is in quite good agreement with the data~\cite{Be05}.      The 
effect  on  the  analyzing  power  is  no  less remarkable with a good 
representation now of the data to $\sim 50^\circ$.         As with the 
phenomenological model analysis~\cite{Be05},        our results do not  
compare with the analyzing power data at larger scattering angles, but 
in that region the cross sections are less than 0.1~mb/sr.    For such 
small values, with elastic scattering,   the scattering model used may 
not suffice. Furthermore, as all components have large negative values 
of analyzing power in the vicinity of $60^\circ$ scattering angle,  to 
match the data of essentially a null result  requires  that individual 
amplitudes destructively interfere.    Small factors can influence the 
phases to effect such.


\subsection{Inelastic scattering of 197 MeV protons}

Betker {\it et al.}~\cite{Be05} show cross-section and spin observable 
data from the inelastic scattering of 197 MeV polarized protons   with 
$^{10}$B and for the same set of states for which     we have analyzed 
electron scattering form factors.  To analyze their data, we have used 
the DWA with distorted wave functions generated from the   $g$-folding 
model of the optical potentials, the Melbourne force  $g_{\text{eff}}$ 
used as the transition operator,  and the nuclear state and transition 
details  being  those  from the  $(0+2)\hbar\omega$  shell  model  for 
$^{10}$B the same as used in the analyses of the electron   scattering 
form factors. Consequently, each result shown hereafter was found with 
but one run of the DWBA98 code~\cite{Ra98}.      No {\it a posteriori} 
adjustments have been made to improve agreement with the data.

\subsubsection{Cross sections and analyzing powers to the $1^+$ states}

\begin{figure}[h]
\begin{center}
\scalebox{0.7}{\includegraphics*{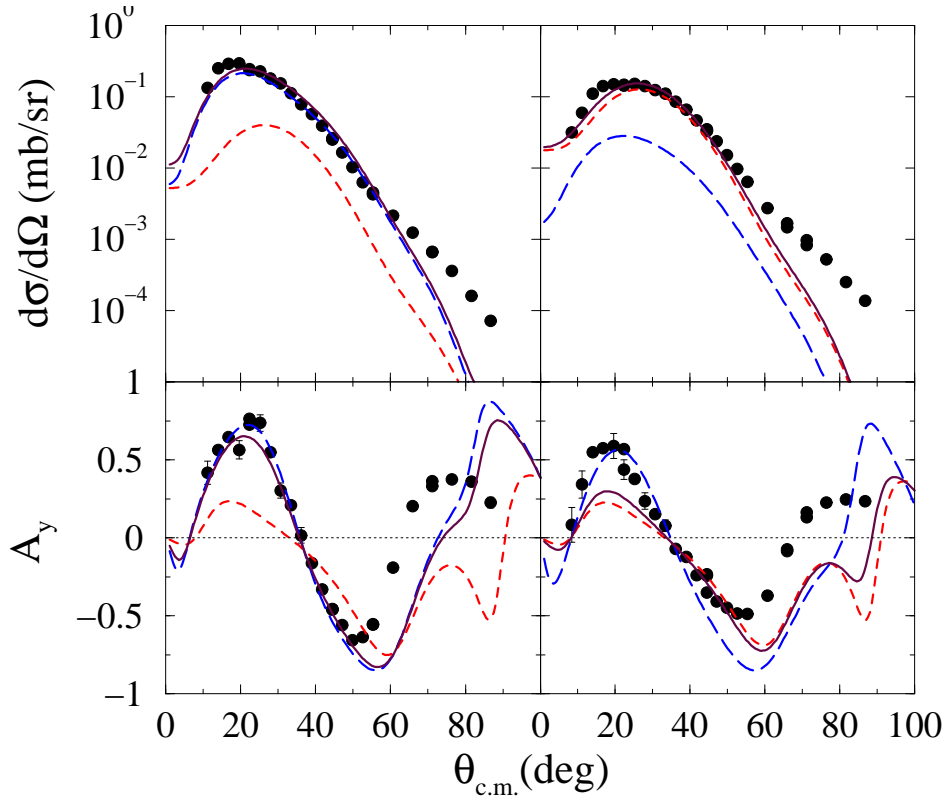}}
\end{center}
\caption{\label{Fig-197-1plus} (Color online)
The inelastic scattering cross section (top) and analyzing
power (bottom) of 197 MeV polarized protons exciting the 
$1^+$ states of $^{10}$B.  The data~\cite{Be05} are compared to  
results as described in the text.}
\end{figure}
In Fig.~\ref{Fig-197-1plus} we show cross sections (top) and analyzing
powers (bottom)  for the excitation of the $1^+$ 0.718 MeV (left)  and 
of the $1^+$ 2.154 MeV (right) states.       The data, depicted by the
filled circles, are compared with the results of  our DWA calculations
with the $I=2$ contributions depicted by the long dashed curves,   the 
$I=3$ ones by the dashed curves, and our complete results by the solid 
curves.

Our cross section results are in quite good agreement  with  the  data
to $\sim 50^\circ$ scattering angle.  Thereafter the calculated values 
decrease more rapidly than observation.          However for the large 
scattering angles the data are less than $\sim 0.001$ mb/sr and are at 
least an order of magnitude smaller than the peak value.         Small 
variations, most notably in the structure of the  single-nucleon  wave 
functions within the nuclear volume could account for that.  The small 
admixture  of  the  shell model  states  that  gave  such  a  dramatic 
improvement in comparisons of the  electron  scattering  form factors, 
does little in these cases. As with the form factors for these states, 
the  proton  scatterings  are  mixes of scattering amplitudes for both 
$I=2$ and $I=3$ angular momentum transfers; the former dominant in the 
$1^+$ 0.718 MeV state excitation,    the latter in the $1^+$ 2.154 MeV 
state case.  Analyzing powers, being normalized by the cross sections,
then also reflect the shape of the dominant components in those  cross 
sections. In comparison with the data, the dominant $I=2$ character of 
the $1^+$ 0.718 MeV state excitation is very evident.

\subsubsection{Excitation of the $3^+$ 4.774 MeV state}

\begin{figure}[h]
\begin{center}
\scalebox{0.65}{\includegraphics*{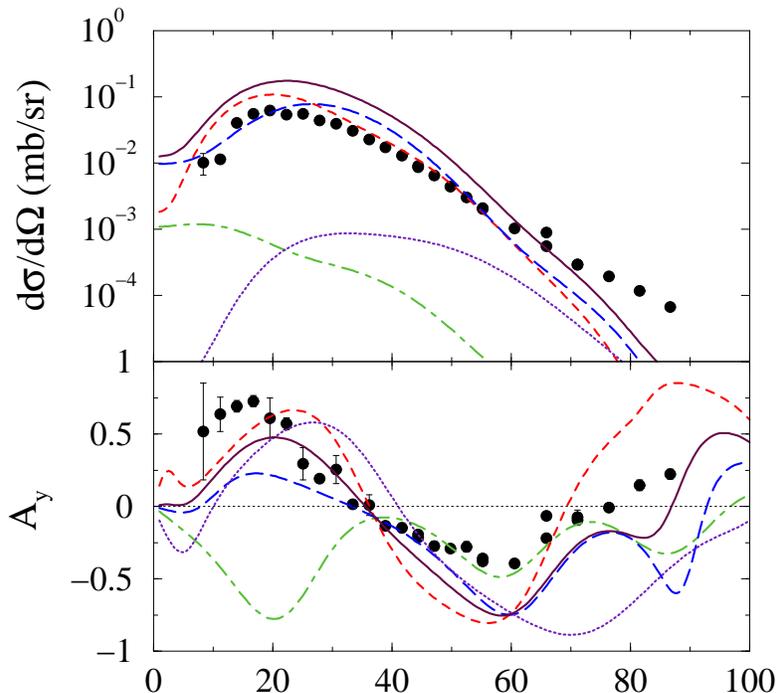}}
\end{center}
\caption{\label{Fig-197-3plus} (Color online)
The inelastic scattering cross section (top) and analyzing
power (bottom) of 197 MeV polarized protons exciting the 
$3^+;$ 4.774 MeV state of $^{10}$B.}
\end{figure}
The cross sections and analyzing powers for  excitation  of the  $3^+$ 
4.774 MeV state in $^{10}$B are shown in the top  and  bottom parts of
Fig.~\ref{Fig-197-3plus} respectively.    The data~\cite{Be05} (filled 
circles) are compared to the  DWA calculated  results  for  purely  an 
$I=2$ angular momentum transfer  (dashed curves),   for a purely $I=3$ 
case (long dashed curves)  and for the full result (solid curves). The 
small $I=1$ and $I=4$ components are also displayed by the  dot-dashed 
and dotted curves respectively. The allowed $I=0$ component is smaller 
than $10^{-4}$ mb/sr and so is not plotted.

The cross section is over  predicted,  as  was the electron scattering 
form factor for this transition,       and thus we need to improve the 
structure model description of this $3^+$ 4.774 MeV state.   The shape 
of the calculated proton scattering cross section, however,    is very 
similar to that of the data. This is quite different to what was found 
for the longitudinal form factor from electron scattering though   the 
limited data of transverse form factor was reproduced.      The latter
though was dominantly a transition due to an M3 multipole.         The
analyzing power results show that the shape of the cross section   has  
some significance as the components do sum to give a reasonable result.

\subsubsection{Excitation of the $2^+$3.587 MeV and of the $4^+$
6.025 MeV states}

\begin{figure}[h]
\begin{center}
\scalebox{0.7}{\includegraphics*{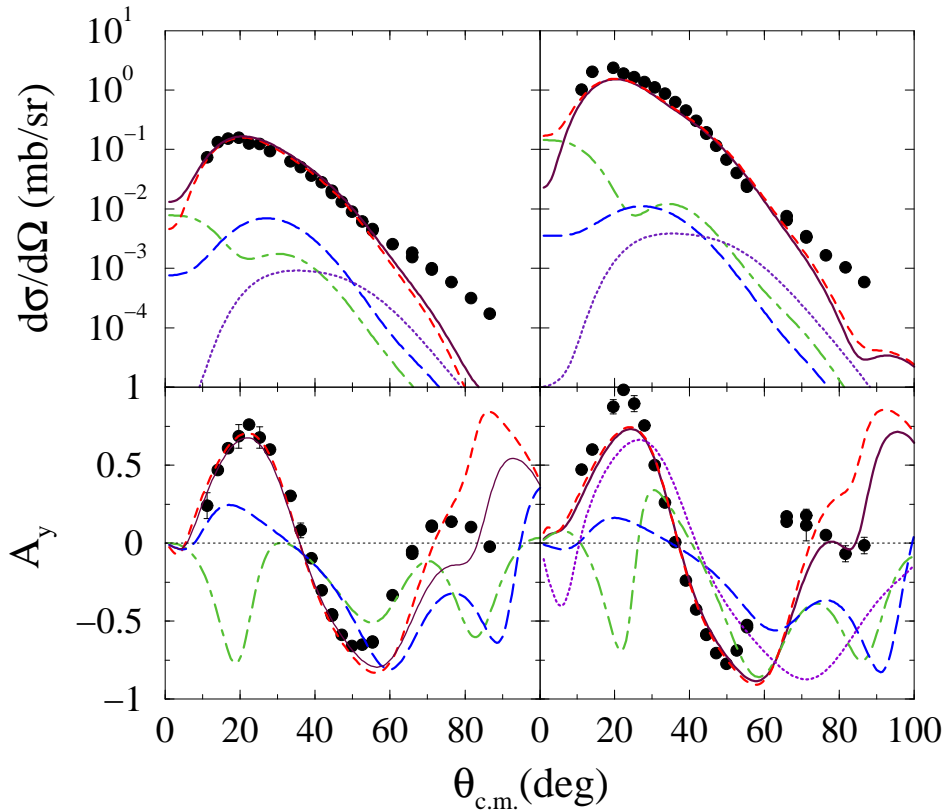}}
\end{center}
\caption{\label{Fig-197-2and4plus} (Color online)
The inelastic scattering cross section (top) and analyzing
power (bottom) of 197 MeV polarized protons exciting the 
$2^+$ 3.587 MeV (left) and $4^+$ 6.025 MeV (right) states in
$^{10}$B. The data~\cite{Be05} (filled circles)
are compared to the evaluations as described in the text.}
\end{figure}

The  inelastic  scattering  cross  sections and analyzing powers  from 
inelastic scattering of 197 MeV polarized protons exciting  the $2^+$ 
3.587~MeV  and  $4^+$ 6.025~MeV states in  $^{10}$B are displayed in 
Fig.~\ref{Fig-197-2and4plus}.
The predictions are in good agreement with data,    to $\sim 60^\circ$ 
for cross sections $\ge 10^{-3}$ mb/sr.          The total results are 
portrayed by the solid lines and those for individual angular momentum 
transfer values are shown by the dot-dashed curves ($I=1$),     by the 
dashed curves ($I=2$), by the long dashed curves ($I=3$),   and by the 
dotted curves ($I=4$). Higher multipole contributions are not shown as 
they have peak values less than $10^{-5}$ mb/sr.  These results mirror 
the fits to  data  that  we  found  with  the chosen structure for the 
electron scattering form factors in that  the  dominant  contributions 
are of quadrupole type and in the quality of fit to cross section data 
to $\sim 3$ fm$^{-1}$ linear momentum transfer.    The analyzing power 
data also are well reproduced at least to $\sim 50^\circ$. 

\subsubsection{Inelastic excitation of the $0^+;T=1$ state
and ($p,n$) scattering to its analogue}

\begin{figure}[h]
\begin{center}
\scalebox{0.7}{\includegraphics*{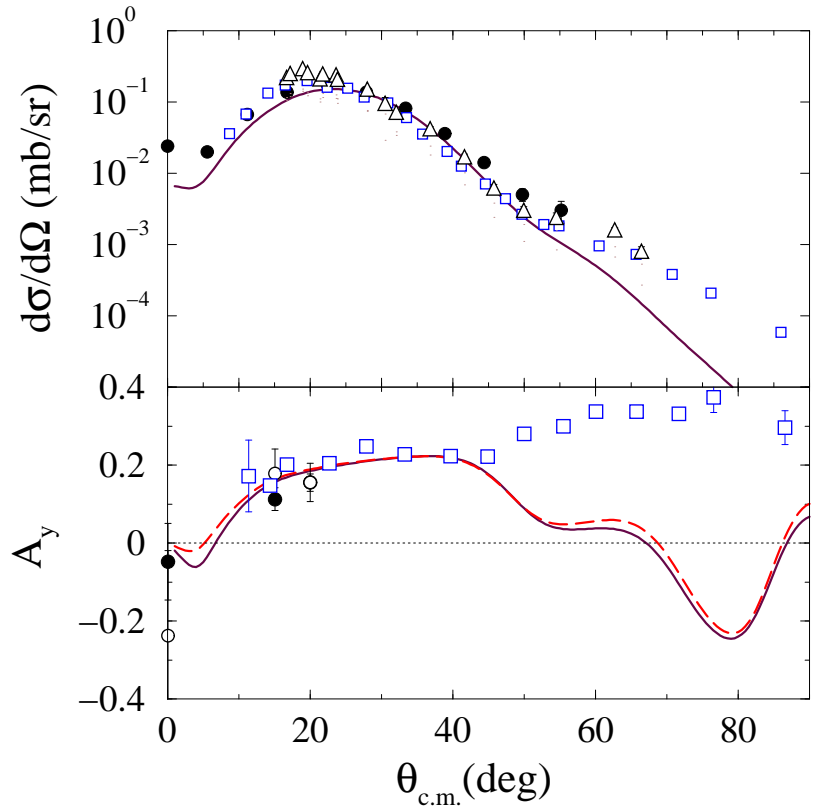}}
\end{center}
\caption{\label{Fig-197-pn} (Color online)
Cross sections (top) and analyzing powers (bottom) from the 186 MeV 
$(p,n)$ reaction to the ground state $(0^+;T=1)$ of $^{10}$C.  
Included are data, with cross section scaled by a factor of 2, from 
the inelastic scattering (of 200 MeV protons) to the isobaric analogue 
state (1.74 MeV) in $^{10}$B.}
\end{figure}

Finally, in Fig.~\ref{Fig-197-pn}, cross sections and analyzing powers 
from the 186 MeV $(p,n)$ reaction to the ground state $(0^+;T=1)$   of 
$^{10}$C and from the inelastic scattering of 200 MeV protons to the 
isobaric analogue state (1.74 MeV) in $^{10}$B are shown.    The OBDME 
for the inelastic scattering to the analogue scale to those    for the 
charge exchange reaction  by  an  isospin  Clebsch-Gordan coefficient. 
Hence  the  cross  sections  for  inelastic scattering to the 1.74 MeV 
$0^+;T=1$ state of $^{10}$B depicted in  Fig.~\ref{Fig-197-pn}  have 
been scaled by a factor of 2. 
In this figure,     the charge exchange data are depicted by the solid 
circles while the data from the inelastic scattering       (of 200 MeV 
protons) to the isobaric analogue  state  are  depicted  by  the  open 
squares.   The solid curves are the results we have obtained using our 
model of structure.  For comparison, the dashed curve in the analyzing 
power segment  is  the  result  we  found for the polarization in this 
(inelastic) transition.        The differences between calculated spin 
measurables are trivial.   In this case we have shown results and data 
taken to a scattering angle of 90$^\circ$,  and  it  is appropriate to 
view them in two parts;         for scattering angles below  and above 
$\sim 50^\circ$.      At that boundary, the cross sections have values 
$\sim 10^{-3}$ mb/sr which is $\sim 1$\% of the peak values.   As with 
our results for the other transitions,        for scattering angles to 
$50^\circ$, the predictions  for both the cross sections and analyzing 
powers are in quite good agreement with data. Likewise, the calculated 
cross sections decrease too rapidly for the larger  scattering angles. 
That mismatch makes meaningless consideration of what results for spin 
observables  given  that  those are normalized by  the  cross  section 
values. However, we stress that it is a small magnitude,        larger 
momentum transfer, character of the processes that are in error.  Such 
can reflect needed improvements in one or more of the internal  radial 
wave functions,         the higher momentum transfer properties of the 
effective $NN$ interaction,   and/or the reaction mechanism specifics.  
But to pay most attention to these deficiencies is to let    ``the tail 
wag the dog''.  We do predict the appropriate structures and magnitudes 
of the data to $\sim 50^\circ$     where the cross sections have their 
largest values.   Consequently the dominant reaction mechanism and its 
details, including the choice of nuclear structure are credible.

\subsection{Other spin observables}

Additional  to  the  analyzing  powers  (and polarizations),  the spin 
observables of polarization transfer coefficients   have been measured
\cite{Be05,Ba92,Wa93}.           The data taken by Betker {\it et al.} 
\cite{Be05} from $^{10}$B (which has a ground state of     spin-parity 
$3^+$),    reasonably satisfy the conditions noted by Liu {\it et al.}
\cite{Li96} and given in Eq.~(\ref{Symcon}).         Those are met for 
scattering from spin zero targets whence there should be    only three 
independent coefficients.    Nonetheless we have evaluated all five to 
see  how  well  those  links  are satisfied by the assumed shell model 
spectroscopy.

\subsubsection{The polarization coefficients, $D_{NN'}$}

\begin{figure}[h]
\begin{center}
\scalebox{0.7}{\includegraphics*{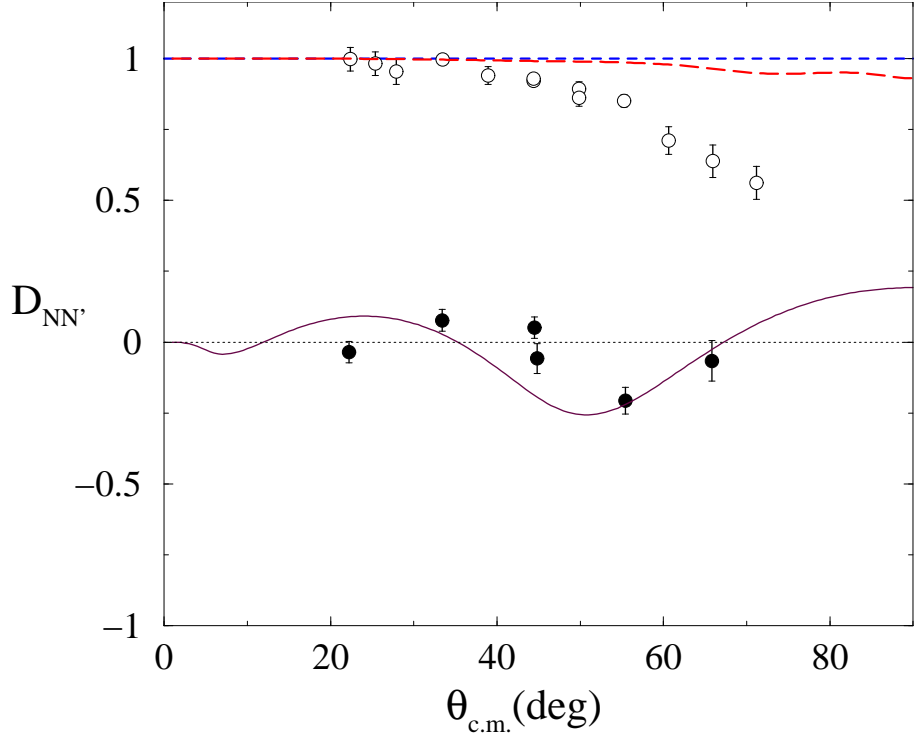}}
\end{center}
\caption{\label{Fig-197-gs+1.74-Dnn} (Color online)
Polarization transfer coefficients $D_{NN'}$ from the elastic 
scattering of 197 MeV protons from $^{10}$B and from inelastic 
scattering to the $(0^+;T=1)$ at 1.74 MeV.}
\end{figure}

In Fig.~\ref{Fig-197-gs+1.74-Dnn} the $D_{NN'}$ from both the  elastic
and inelastic scattering (to the $(0^+;T=1)$ 1.74 MeV state) of    197 
MeV protons are shown.     The elastic scattering data~\cite{Be05} are
depicted by the opaque, and the inelastic scattering  data~\cite{Ba92} 
by the filled circles respectively.
The elastic scattering data are compared with calculated results   for 
$I=0$ (dashed curve) and $I=2$ (long dashed curve) elastic  scattering 
components.   The inelastic result involves the unique $I=3$ transfer.
These angular  momentum  transfer components dominate the transitions.
The elastic scattering data are close to the value 1,   as expected by
the symmetry condition in Eq.~\ref{Symcon}, up to $\sim 40^\circ$.  At 
larger scattering angles they decrease significantly. Our two results,
for the pure multipoles, $D_{NN'}^{(2)}$ and $D_{NN'}^{(3)}$,  as well 
as the total that can be formed by the summation,
\begin{equation}
D_{NN'} = \frac{\sigma_2 D_{NN'}^{(2)} + \sigma_3 D_{NN'}^{(3)}}
{[\sigma_{2} + \sigma_{3}]}\ ,
\end{equation}
where $\sigma_{I}$ are the pure multipole ($I$)  differential    cross 
sections, deviate far less from 1 than the data.   This effect is very 
similar to that found by Betker {\it et al.}~\cite{Be05}.

The  $D_{NN'}$  for the unnatural parity transition to the  $0^+; T=1$ 
1.74 MeV state lies close to zero and our (pure $I=3$)  result  agrees 
quite well with the data~\cite{Ba92}  shown  by  the  filled  circles. 
Though our effective  interaction  is  a complicated  mix  of operator 
terms,  it  is  preset  for  all  calculations  of all transitions and 
observables. It is medium dependent and so the improvement in results, 
noted~\cite{Ba92}  by  use  of effective-mass  approximations  upon  a 
phenomenological Franey-Love interaction,   is  confirmed  as well  as 
improved.

\begin{figure}[h]
\begin{center}
\scalebox{0.7}{\includegraphics*{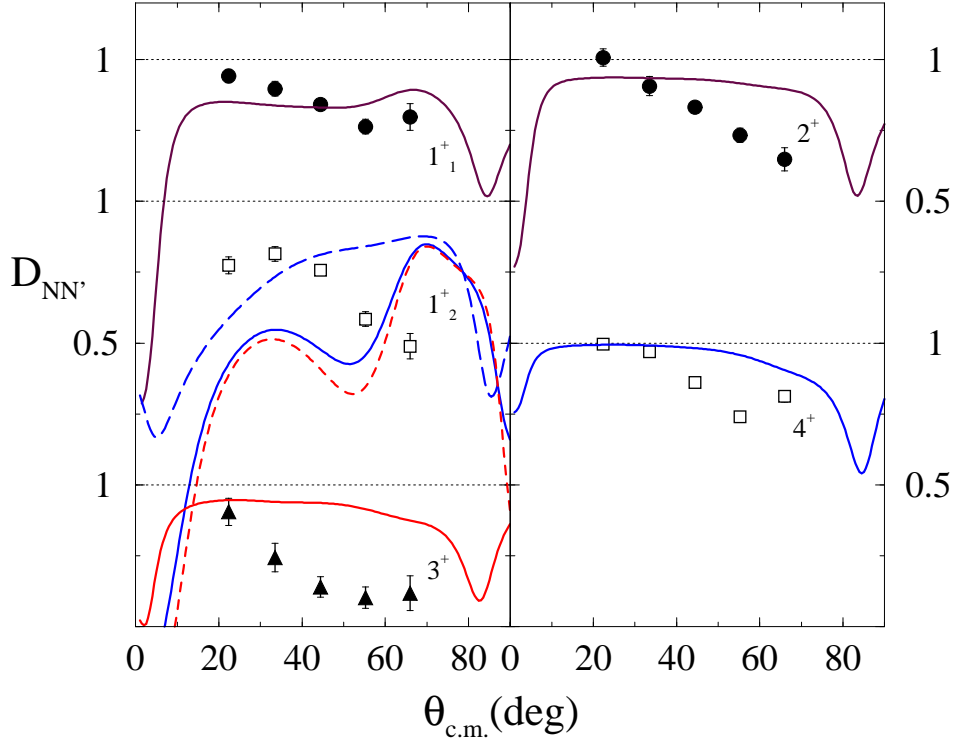}}
\end{center}
\caption{\label{B10-197-other-Dnn} (Color online)
Polarization transfer coefficients $D_{NN}$ 
from inelastic scattering of 197 MeV protons from ${}^{10}$B
leading to the states identified by their values of $J^\pi$.
}
\end{figure}

The $D_{NN'}$ measured with the inelastic scattering to other states in
${}^{10}$B  are  compared  with  the  results  of  our  calculations in
Fig.~\ref{B10-197-other-Dnn}.
In this figure,           the results shown in the left panel are from
excitation of the $1^+$ 0.718 MeV state (top),  from the excitation of
the $1^+$ 2.15 MeV state (middle),      and from the excitation of the 
$3^+$ 4.774 MeV state (bottom).     In the right panel the results for 
excitation of the $2^+$ 3.59 MeV and of the  $4^+$ 6.02 MeV states are 
depicted in the top and bottom parts respectively.    The solid curves 
display the complete results when both $I=2$ and $I=3$   contributions 
dominate.     The excitations of the $2^+$ and $4^+$ states are almost 
pure $I=2$.    With the $1^+_2$ (2.15 MeV) results we also display the
separate $I=2$  (long dashed curve)  and $I=3$ (dashed curve) results. 
Clearly in this case the octupole is the more significant term.   With 
the   $1^+_1$ (0.718 MeV) transition  the  $I=2$ component is the more 
significant. 

With the exception of our results for the $1^+_2$ excitation,    these 
evaluated $D_{NN'}$ resemble those found by        Betker {\it et al.}
\cite{Be05}. The mix of $I=2$ and $I=3$ components in our  shell model 
structure gives a good result  but it does not do well for the $1^+_2$ 
transition.    The small admixtures favored by the electron scattering 
form factor results little alters these findings.    The comparison of 
results and data for the $3^+$ state excitation also is poor though it 
is  to  be  remembered that the cross-section data for this transition
also were overestimated.       There is clearly a need for an improved 
description of the $3^+$ state.

\subsubsection{The polarization transfer coefficients, 
$D_{LL'}$, $D_{SS'}$, $D_{LS'}$, $D_{SL'}$.}

\begin{figure}[h]
\begin{center}
\scalebox{0.7}{\includegraphics*{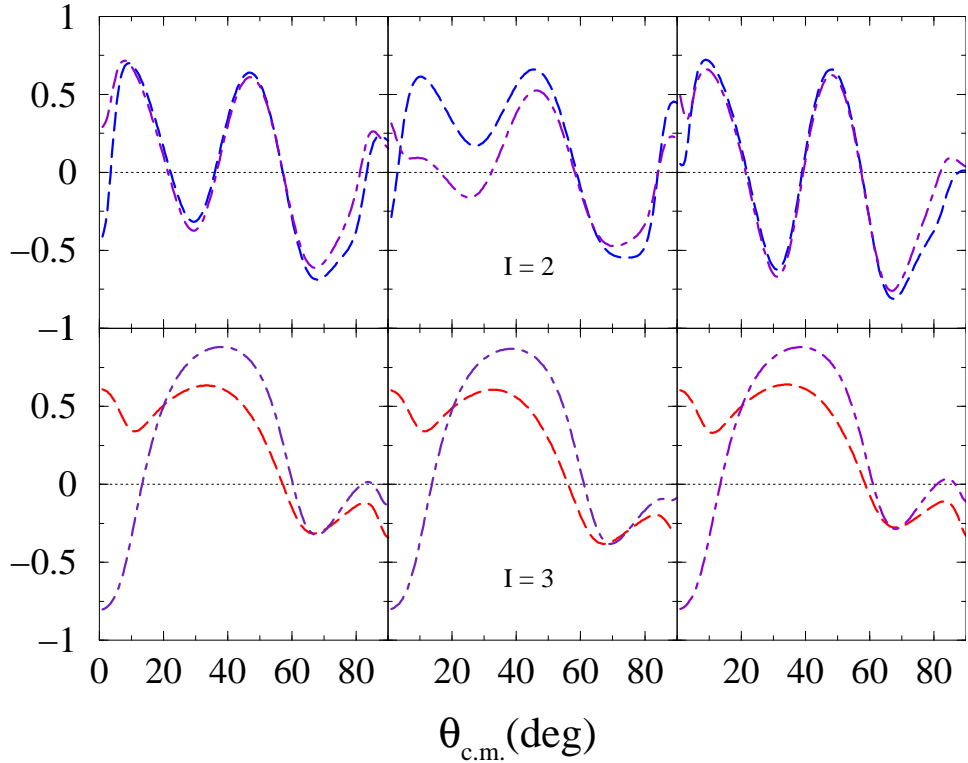}}
\end{center}
\caption{\label{B10-197-1+1+3+-DllDss} (Color online)
Polarization transfer coefficients $D_{LL'}$ (long dashed curves)
and $D_{SS'}$ (dot-dashed curves) for inelastic scattering of 197 MeV 
protons from ${}^{10}$B exciting the $1^+$ 0.718 MeV (left), $1^+$ 
2.154 MeV (middle), and $3^+$ 4.774 MeV (right) states.}
\end{figure}

The  dominant  component contributions to the calculated coefficients,
$D_{LL'}$,       and       $D_{SS'}$       are       displayed      in 
Fig.~\ref{B10-197-1+1+3+-DllDss}.
The top line of these diagrams contain  the  contributions for angular 
momentum transfer $I=2$ while the bottom line shows the results    for 
$I=3$.  From left to right the results are those for excitation of the 
$1^+_1$, the $1^+_2$, and the $3^+$ states respectively. As evident in 
the top panels, the $I=2$ components are very similar and largely   in 
line with the symmetry conditions of Eq.~(\ref{Symcon}).     There are 
some divergence between the $D_{LL'}$ and $D_{SS'}$     with the $I=3$ 
components (shown in the bottom segments), however.        Both have a 
positive   peak   at   $\sim 40^\circ$,   and  a close agreement above 
$\sim 60^\circ$,       but at forward scattering angles they are quite 
different.

\begin{figure}[h]
\begin{center}
\scalebox{0.7}{\includegraphics*{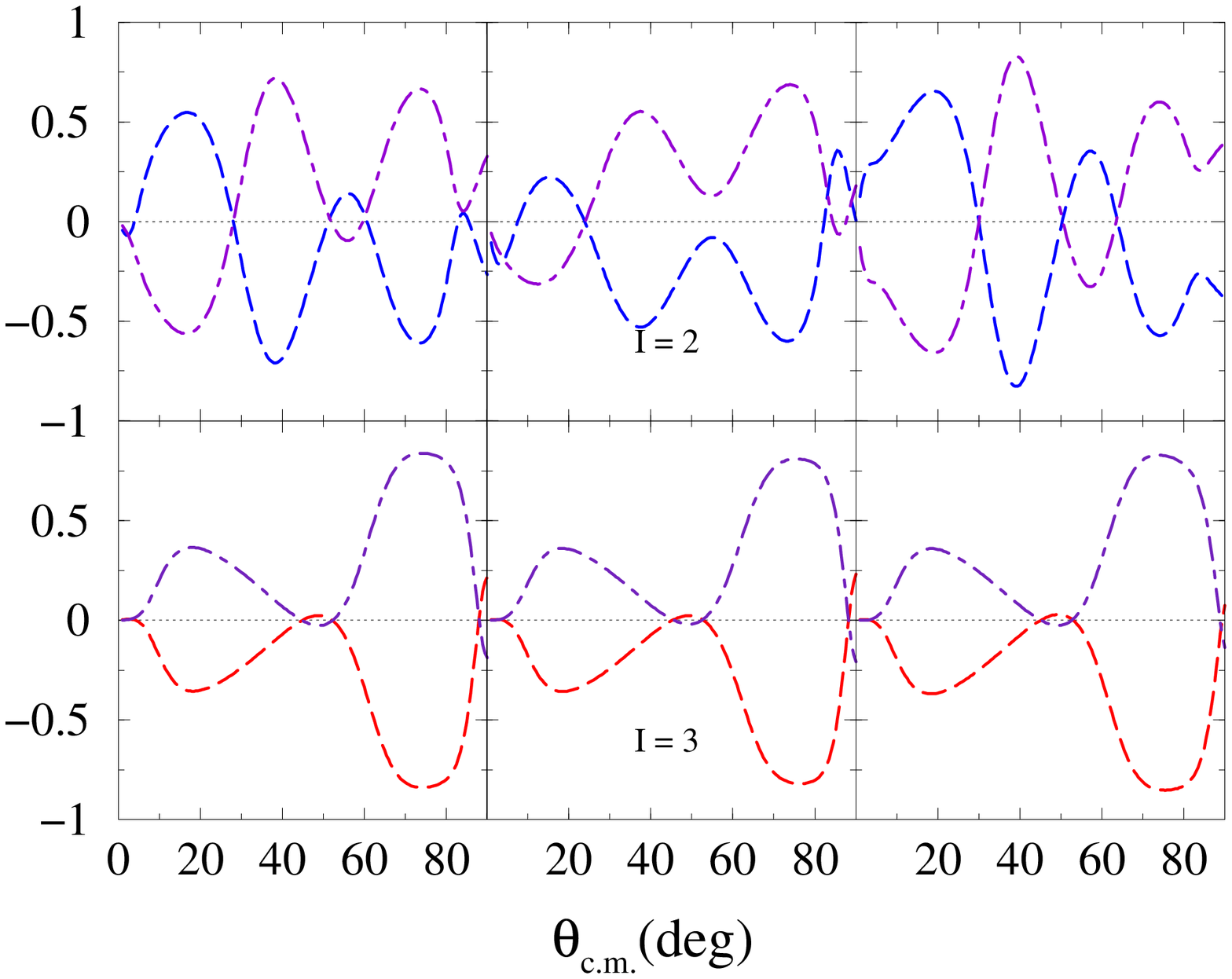}}
\end{center}
\caption{\label{B10-197-1+1+3+-DlsDsl} (Color online)
Polarization transfer coefficients $D_{SL'}$ (long dashed curves)
and $D_{LS'}$ (dot-dashed curves) for inelastic scattering of 197 MeV 
protons from ${}^{10}$B exciting the $1^+$ 0.718 MeV (left), $1^+$ 
2.154 MeV (middle), and $3^+$ 4.774 MeV (right) states.}
\end{figure}

For  the same three states,            and  with specifics as given in 
Fig.~\ref{B10-197-1+1+3+-DllDss},         the component values of  the
polarization transfer observables $D_{LS'}$ and $D_{SL'}$ are depicted 
in Fig.~\ref{B10-197-1+1+3+-DlsDsl}
As with the $D_{NN'}$ and $D_{SS'}$,    the $I=2$ and $I=3$ components 
for each state are markedly different in shape. However, for all three 
transitions the results are similar ($I=2$)   and very similar ($I=3$) 
with  differences  among the latter  being no more than a few percent.  
With these observables, and for both angular momentum transfer values,
the mirror symmetry of the conditions of Eq.~(\ref{Symcon}) is  clear.

\begin{figure}[h]
\begin{center}
\scalebox{0.7}{\includegraphics*{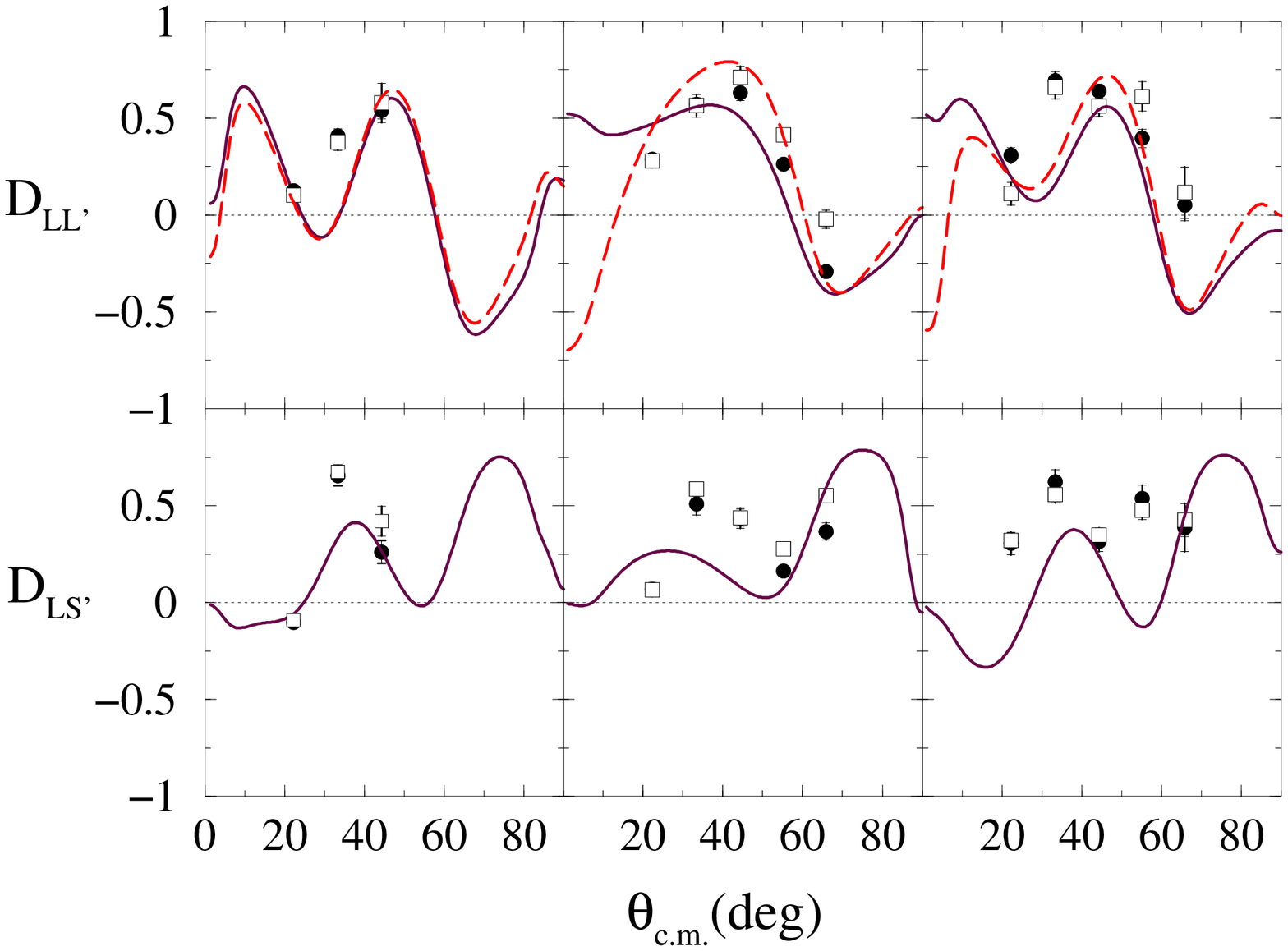}}
\end{center}
\caption{\label{B10-197-1+1+3+-DllDls} (Color online)
Polarization transfer coefficients for inelastic scattering of 197 MeV 
protons from ${}^{10}$B exciting the $1^+$ 0.718 MeV (left), $1^+$ 
2.154 MeV (middle), and $3^+$ 4.774 MeV (right) states. Other details 
are given in the text.}
\end{figure}

Then, in Fig.~\ref{B10-197-1+1+3+-DllDls},  the complete results   for
the spin observables $D_{LL'}, D_{SS'}$, and $D_{LS'}$    are compared
with data~\cite{Be05}.  
In the top segments,   the solid and dashed curves show the calculated
results for $D_{LL'}$ and $D_{SS'}$ respectively.   The $D_{LL'}$ data 
are displayed by the solid circles while those for $D_{SS'}$ are given 
by the open squares.      The solid curves in the figures shown in the 
bottom segments of this figure are the calculated $D_{LS'}$    results 
(nearly identical to those for $-D_{SL'}$).     They are compared with
data~\cite{Be05}  where the solid circles are the values for $D_{LS'}$ 
and the open squares for $-D_{SL'}$.  Given that spin measurables are 
quite sensitive to details in the evaluations,     these results agree 
quite well with the data from the three transitions.

\begin{figure}[h]
\begin{center}
\scalebox{0.7}{\includegraphics*{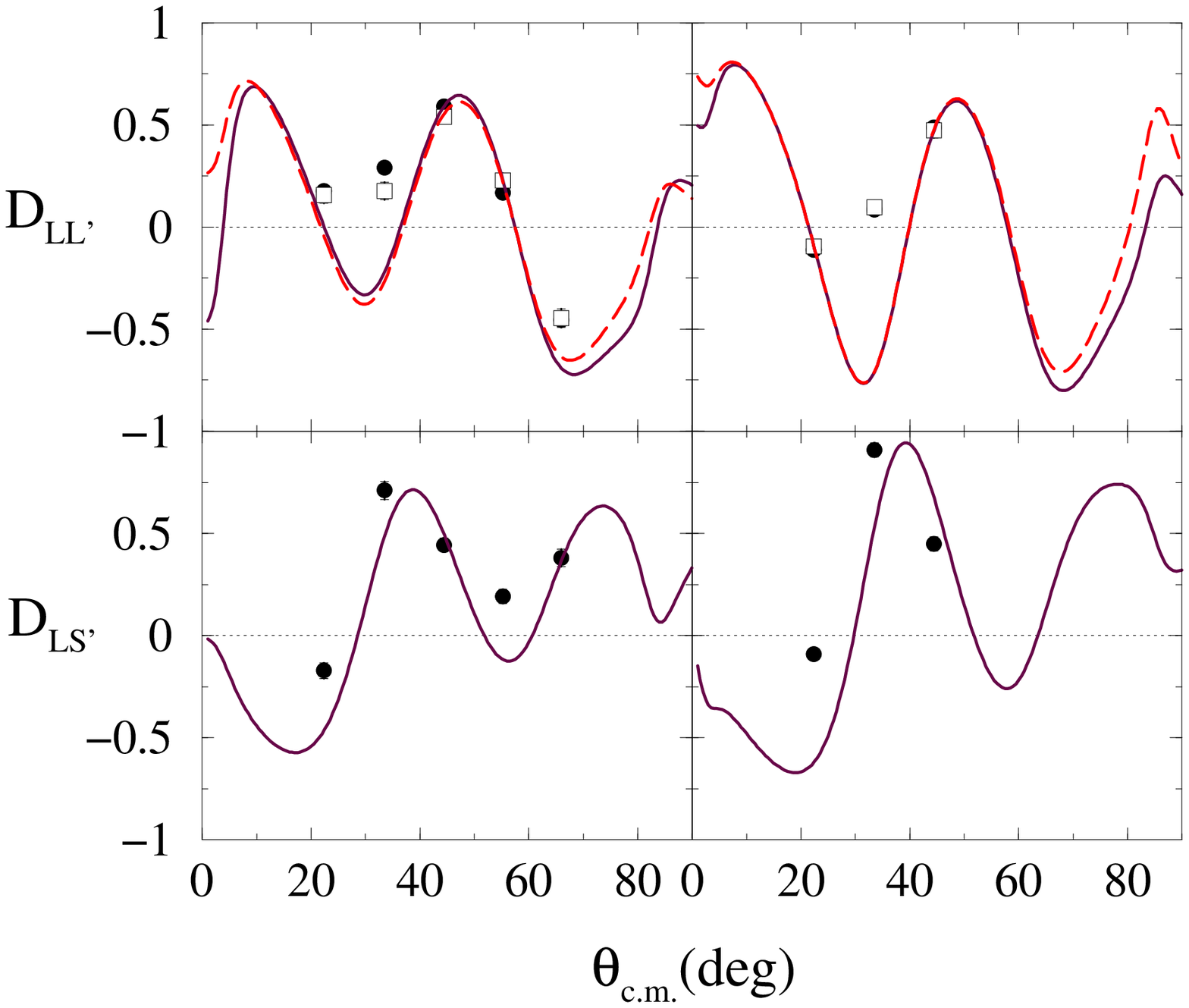}}
\end{center}
\caption{\label{B10-197-2+4+-DllDls} (Color online)
Polarization transfer coefficients, $D_{LL'}$ (top)  and $D_{LS'}$
bottom, for inelastic scattering of 197 MeV protons 
from $^{10}$B exciting the $2^+$ 3.59 MeV 
and the $4^+$ 6.02 MeV states. The results for those two states are 
shown in the left and right panels respectively.}
\end{figure}

Finally,   in Fig.~\ref{B10-197-2+4+-DllDls}, we show the polarization 
transfer coefficients,  $D_{LL'}$ (top)   and $D_{LS'}$ for  inelastic 
scattering of 197 MeV protons from $^{10}$B exciting the    $2^+$ 3.59 
MeV and the $4^+$ 6.02 MeV states.
These transitions are dominated by the   $I=2$   contributions and our
results (solid curves) agree very well with the limited data available. 
For comparison we show by the dashed curves,       our results for the
$D_{SS'}$ coefficients. In both transitions the symmetry condition  of
Eq.~(\ref{Symcon}) is well met.


\section{Conclusions}
\label{Conclusions}

We have made a comprehensive assessment of the structure of the ground
and low excitation states in $^{10}$B.      We have assumed that those
states are described by wave functions determined from a no-core shell
model calculation.        The no-core shell model was defined within a 
complete $(0+2)\hbar\omega$ single particle space and the MK3W   shell 
model interaction used to specify the Hamiltonian.     A most striking
feature of the results of that calculation     was that the ground and 
first excited states are inverted. However, the splitting is but a few 
hundred keV, and with the ``right'' three-nucleon force that inversion 
was effected.      Likewise static moments improved with those quantum 
Monte-Carlo calculations.

Nonetheless,   we persist with the no-core shell model since only with 
its wave functions could we specify OBDME for use in fully microscopic
model studies of electron and medium energy    ($\sim 200$ MeV) proton
scattering.      The scattering calculations are predictions since all
details  required   are   predetermined   in  the  relevant  programs. 
Allowance for  meson   exchange current effects in electron scattering 
was  made  in  the   transverse  electric  form factors by recourse to 
Siegert's theorem.  The proton scattering calculations were made using 
a $g$-folding model for the  optical  potentials  and a  DWA  for  the 
inelastic processes.              The electron scattering form factors  
(longitudinal, electric transverse, and magnetic transverse)     found 
with the no-core shell model wave functions  were  in  good  agreement 
with observation requiring only small admixtures between the two shell 
model $1^+$ states in those transitions. Those $1^+$-state excitations 
however are especially sensitive as the neutron and proton  amplitudes 
destructively interfere. 

The shell model details were used to predict many observables from the 
scattering of 197 MeV polarized protons from $^{10}$B, for which there 
are much data. The elastic scattering observables were evaluated using 
the nonlocal optical potential formed by $g$-folding of the model wave 
functions with a complex, medium-dependent, effective $NN$ interaction. 
The resultant cross section matched data well     as did the analyzing 
power,      at least to momentum transfers for which the cross section 
exceeded $\sim 0.1$ mb/sr in size.  Likewise on using a DWA, the cross 
sections from inelastic scattering to the low excitation states   also 
matched data quite well at momentum transfer values for which    those 
cross sections exceed $\sim 10^{-3}$ mb/sr.   By and large so did spin 
observable results. 

The quality of match between our predictions of these    complementary 
scattering data  involving  the  ground  and  low excitation states in 
$^{10}$B, is evidence that the wave functions for the nucleus obtained 
from   the   $(0+2)\hbar\omega$  shell  model  are  quite   reasonable 
descriptions of those states. The slight disparities in the calculated 
spectrum compared with the known one at low excitation  (a few hundred 
keV) are due to missing elements in the shell model Hamiltonian,  such 
as three-nucleon force effects.   But their inclusion in the structure 
calculation should not seriously alter the eigenvectors.  At best, the 
inclusion of such terms in the  Hamiltonian  should  only give rise to 
small perturbations in the wave functions.

\begin{acknowledgments}
This research was supported  by  a  2007  research grant of the  Cheju 
National University.  The research was also supported by the  National 
Research Foundation, South Africa.
\end{acknowledgments}

\bibliography{B10-paper}

\begin{thebibliography}{30}
\expandafter\ifx\csname natexlab\endcsname\relax\def\natexlab#1{#1}\fi
\expandafter\ifx\csname bibnamefont\endcsname\relax
  \def\bibnamefont#1{#1}\fi
\expandafter\ifx\csname bibfnamefont\endcsname\relax
  \def\bibfnamefont#1{#1}\fi
\expandafter\ifx\csname citenamefont\endcsname\relax
  \def\citenamefont#1{#1}\fi
\expandafter\ifx\csname url\endcsname\relax
  \def\url#1{\texttt{#1}}\fi
\expandafter\ifx\csname urlprefix\endcsname\relax\def\urlprefix{URL }\fi
\providecommand{\bibinfo}[2]{#2}
\providecommand{\eprint}[2][]{\url{#2}}

\bibitem[{\citenamefont{Betker et~al.}(2005)}]{Be05}
\bibinfo{author}{\bibfnamefont{A.~C.} \bibnamefont{Betker}}
  \bibnamefont{et~al.}, \bibinfo{journal}{Phys. Rev. C}
  \textbf{\bibinfo{volume}{71}}, \bibinfo{pages}{064607}
  (\bibinfo{year}{2005}).

\bibitem[{\citenamefont{Wang et~al.}(1993)}]{Wa93}
\bibinfo{author}{\bibfnamefont{L.}~\bibnamefont{Wang}} \bibnamefont{et~al.},
  \bibinfo{journal}{Phys. Rev. C} \textbf{\bibinfo{volume}{47}},
  \bibinfo{pages}{2123} (\bibinfo{year}{1993}).

\bibitem[{\citenamefont{Cichocki et~al.}(1995)\citenamefont{Cichocki, Dubach,
  Hicks, Peterson, de~Jaeger, de~Vries, Kalantar-Nayestanaki, and Sato}}]{Ci95}
\bibinfo{author}{\bibfnamefont{A.}~\bibnamefont{Cichocki}},
  \bibinfo{author}{\bibfnamefont{J.}~\bibnamefont{Dubach}},
  \bibinfo{author}{\bibfnamefont{R.~S.} \bibnamefont{Hicks}},
  \bibinfo{author}{\bibfnamefont{G.~A.} \bibnamefont{Peterson}},
  \bibinfo{author}{\bibfnamefont{C.~W.} \bibnamefont{de~Jaeger}},
  \bibinfo{author}{\bibfnamefont{H.}~\bibnamefont{de~Vries}},
  \bibinfo{author}{\bibfnamefont{N.}~\bibnamefont{Kalantar-Nayestanaki}},
  \bibnamefont{and} \bibinfo{author}{\bibfnamefont{T.}~\bibnamefont{Sato}},
  \bibinfo{journal}{Phys. Rev. C} \textbf{\bibinfo{volume}{51}},
  \bibinfo{pages}{2406} (\bibinfo{year}{1995}).

\bibitem[{\citenamefont{Amos et~al.}(2000)\citenamefont{Amos, Dortmans, von
  Geramb, Karataglidis, and Raynal}}]{Am00}
\bibinfo{author}{\bibfnamefont{K.}~\bibnamefont{Amos}},
  \bibinfo{author}{\bibfnamefont{P.~J.} \bibnamefont{Dortmans}},
  \bibinfo{author}{\bibfnamefont{H.~V.} \bibnamefont{von Geramb}},
  \bibinfo{author}{\bibfnamefont{S.}~\bibnamefont{Karataglidis}},
  \bibnamefont{and} \bibinfo{author}{\bibfnamefont{J.}~\bibnamefont{Raynal}},
  \bibinfo{journal}{Adv. in Nucl. Phys.} \textbf{\bibinfo{volume}{25}},
  \bibinfo{pages}{275} (\bibinfo{year}{2000}), \bibinfo{note}{(and references
  contained therein)}.

\bibitem[{\citenamefont{Karataglidis et~al.}(1995)\citenamefont{Karataglidis,
  Halse, and Amos}}]{Ka95}
\bibinfo{author}{\bibfnamefont{S.}~\bibnamefont{Karataglidis}},
  \bibinfo{author}{\bibfnamefont{P.}~\bibnamefont{Halse}}, \bibnamefont{and}
  \bibinfo{author}{\bibfnamefont{K.}~\bibnamefont{Amos}},
  \bibinfo{journal}{Phys. Rev. C} \textbf{\bibinfo{volume}{51}},
  \bibinfo{pages}{2494} (\bibinfo{year}{1995}).

\bibitem[{\citenamefont{Warburton and Millener}(1989)}]{Wa89}
\bibinfo{author}{\bibfnamefont{E.~K.} \bibnamefont{Warburton}}
  \bibnamefont{and} \bibinfo{author}{\bibfnamefont{D.~J.}
  \bibnamefont{Millener}}, \bibinfo{journal}{Phys. Rev. C}
  \textbf{\bibinfo{volume}{39}}, \bibinfo{pages}{1120} (\bibinfo{year}{1989}).

\bibitem[{\citenamefont{Hannen et~al.}(2003)}]{Ha03}
\bibinfo{author}{\bibfnamefont{V.~M.} \bibnamefont{Hannen}}
  \bibnamefont{et~al.}, \bibinfo{journal}{Phys. Rev. C}
  \textbf{\bibinfo{volume}{67}}, \bibinfo{pages}{054230, 054231}
  (\bibinfo{year}{2003}), \bibinfo{note}{(and references contained therein)}.

\bibitem[{\citenamefont{Raynal}(1998)}]{Ra98}
\bibinfo{author}{\bibfnamefont{J.}~\bibnamefont{Raynal}}
  (\bibinfo{year}{1998}), \bibinfo{note}{computer program DWBA98, NEA 1209/05}.

\bibitem[{\citenamefont{Amos et~al.}(2003)\citenamefont{Amos, Canton, Pisent,
  Svenne, and van~der Knijff}}]{Am03}
\bibinfo{author}{\bibfnamefont{K.}~\bibnamefont{Amos}},
  \bibinfo{author}{\bibfnamefont{L.}~\bibnamefont{Canton}},
  \bibinfo{author}{\bibfnamefont{G.}~\bibnamefont{Pisent}},
  \bibinfo{author}{\bibfnamefont{J.~P.} \bibnamefont{Svenne}},
  \bibnamefont{and} \bibinfo{author}{\bibfnamefont{D.}~\bibnamefont{van~der
  Knijff}}, \bibinfo{journal}{Nucl.\ Phys.} \textbf{\bibinfo{volume}{A728}},
  \bibinfo{pages}{65} (\bibinfo{year}{2003}).

\bibitem[{\citenamefont{Canton et~al.}(2006)\citenamefont{Canton, Pisent,
  Svenne, Amos, and Karataglidis}}]{Ca06}
\bibinfo{author}{\bibfnamefont{L.}~\bibnamefont{Canton}},
  \bibinfo{author}{\bibfnamefont{G.}~\bibnamefont{Pisent}},
  \bibinfo{author}{\bibfnamefont{J.~P.} \bibnamefont{Svenne}},
  \bibinfo{author}{\bibfnamefont{K.}~\bibnamefont{Amos}}, \bibnamefont{and}
  \bibinfo{author}{\bibfnamefont{S.}~\bibnamefont{Karataglidis}},
  \bibinfo{journal}{Phys. Rev. Lett.} \textbf{\bibinfo{volume}{96}},
  \bibinfo{pages}{072502} (\bibinfo{year}{2006}).

\bibitem[{\citenamefont{von Geramb et~al.}(1975)\citenamefont{von Geramb, Amos,
  Sprickman, Kn{\"o}pfle, Rogge, Ingham, and Mayer-B{\"o}ricke}}]{Ge75}
\bibinfo{author}{\bibfnamefont{H.~V.} \bibnamefont{von Geramb}},
  \bibinfo{author}{\bibfnamefont{K.}~\bibnamefont{Amos}},
  \bibinfo{author}{\bibfnamefont{R.}~\bibnamefont{Sprickman}},
  \bibinfo{author}{\bibfnamefont{K.~T.} \bibnamefont{Kn{\"o}pfle}},
  \bibinfo{author}{\bibfnamefont{M.}~\bibnamefont{Rogge}},
  \bibinfo{author}{\bibfnamefont{D.}~\bibnamefont{Ingham}}, \bibnamefont{and}
  \bibinfo{author}{\bibfnamefont{C.}~\bibnamefont{Mayer-B{\"o}ricke}},
  \bibinfo{journal}{Phys. Rev. C} \textbf{\bibinfo{volume}{12}},
  \bibinfo{pages}{1697} (\bibinfo{year}{1975}).

\bibitem[{\citenamefont{Cohen and Kurath}(1965)}]{Co65}
\bibinfo{author}{\bibfnamefont{S.}~\bibnamefont{Cohen}} \bibnamefont{and}
  \bibinfo{author}{\bibfnamefont{D.}~\bibnamefont{Kurath}},
  \bibinfo{journal}{Nucl. Phys.} \textbf{\bibinfo{volume}{73}},
  \bibinfo{pages}{1} (\bibinfo{year}{1965}).

\bibitem[{\citenamefont{Millener and Kurath}(1975)}]{Mi75}
\bibinfo{author}{\bibfnamefont{D.~J.} \bibnamefont{Millener}} \bibnamefont{and}
  \bibinfo{author}{\bibfnamefont{D.}~\bibnamefont{Kurath}},
  \bibinfo{journal}{Nucl. Phys.} \textbf{\bibinfo{volume}{A255}},
  \bibinfo{pages}{315} (\bibinfo{year}{1975}), \bibinfo{note}{and references
  cited therein}.

\bibitem[{\citenamefont{Brown et~al.}(1986)\citenamefont{Brown, Etchegoyen, and
  Rae}}]{Br86}
\bibinfo{author}{\bibfnamefont{B.~A.} \bibnamefont{Brown}},
  \bibinfo{author}{\bibfnamefont{A.}~\bibnamefont{Etchegoyen}},
  \bibnamefont{and} \bibinfo{author}{\bibfnamefont{W.~D.~M.}
  \bibnamefont{Rae}}, \emph{\bibinfo{title}{Msucl report no. 524
  (unpublished)}} (\bibinfo{year}{1986}), \bibinfo{note}{OXBASH (the
  Oxford-Buenos-Aries-Michigan-State University shell model code), A
  Etchegoyen, W. D. M. Rae, and N. S. Godwin (MSU version by B. A. Brown,
  1986)}.

\bibitem[{\citenamefont{Tilley et~al.}(2004)\citenamefont{Tilley, Kelley,
  Godwin, Millener, Purcell, Sheu, and Weller}}]{TUNL04}
\bibinfo{author}{\bibfnamefont{D.~R.} \bibnamefont{Tilley}},
  \bibinfo{author}{\bibfnamefont{J.~H.} \bibnamefont{Kelley}},
  \bibinfo{author}{\bibfnamefont{J.~L.} \bibnamefont{Godwin}},
  \bibinfo{author}{\bibfnamefont{D.~J.} \bibnamefont{Millener}},
  \bibinfo{author}{\bibfnamefont{J.~E.} \bibnamefont{Purcell}},
  \bibinfo{author}{\bibfnamefont{C.~G.} \bibnamefont{Sheu}}, \bibnamefont{and}
  \bibinfo{author}{\bibfnamefont{H.~R.} \bibnamefont{Weller}},
  \bibinfo{journal}{Nucl. Phys.} \textbf{\bibinfo{volume}{A745}},
  \bibinfo{pages}{155} (\bibinfo{year}{2004}).

\bibitem[{\citenamefont{Caurier et~al.}(2002)\citenamefont{Caurier,
  Navr\'{a}til, Ormand, and Vary}}]{Ca02}
\bibinfo{author}{\bibfnamefont{E.}~\bibnamefont{Caurier}},
  \bibinfo{author}{\bibfnamefont{P.}~\bibnamefont{Navr\'{a}til}},
  \bibinfo{author}{\bibfnamefont{W.~E.} \bibnamefont{Ormand}},
  \bibnamefont{and} \bibinfo{author}{\bibfnamefont{J.~P.} \bibnamefont{Vary}},
  \bibinfo{journal}{Phys. Rev. C} \textbf{\bibinfo{volume}{66}},
  \bibinfo{pages}{024314} (\bibinfo{year}{2002}).

\bibitem[{\citenamefont{Pieper et~al.}(2002)\citenamefont{Pieper, Varga, and
  Wiringa}}]{Pi02}
\bibinfo{author}{\bibfnamefont{S.~C.} \bibnamefont{Pieper}},
  \bibinfo{author}{\bibfnamefont{K.}~\bibnamefont{Varga}}, \bibnamefont{and}
  \bibinfo{author}{\bibfnamefont{R.~B.} \bibnamefont{Wiringa}},
  \bibinfo{journal}{Phys. Rev. C} \textbf{\bibinfo{volume}{66}},
  \bibinfo{pages}{044310} (\bibinfo{year}{2002}).

\bibitem[{\citenamefont{Haxton and Johnson}(1990)}]{Ha90}
\bibinfo{author}{\bibfnamefont{W.~C.} \bibnamefont{Haxton}} \bibnamefont{and}
  \bibinfo{author}{\bibfnamefont{C.}~\bibnamefont{Johnson}},
  \bibinfo{journal}{Phys. Rev. Lett} \textbf{\bibinfo{volume}{65}},
  \bibinfo{pages}{1325} (\bibinfo{year}{1990}).

\bibitem[{\citenamefont{Ajzenburg-Selove}(1988)}]{Aj88}
\bibinfo{author}{\bibfnamefont{F.}~\bibnamefont{Ajzenburg-Selove}},
  \bibinfo{journal}{Nucl. Phys.} \textbf{\bibinfo{volume}{A490}},
  \bibinfo{pages}{1} (\bibinfo{year}{1988}).

\bibitem[{\citenamefont{Karataglidis and Amos}(2008)}]{Ka08}
\bibinfo{author}{\bibfnamefont{S.}~\bibnamefont{Karataglidis}}
  \bibnamefont{and} \bibinfo{author}{\bibfnamefont{K.}~\bibnamefont{Amos}},
  \bibinfo{journal}{Phys. Lett} \textbf{\bibinfo{volume}{B660}},
  \bibinfo{pages}{428} (\bibinfo{year}{2008}).

\bibitem[{\citenamefont{deForest and Walecka}(1966)}]{Fo66}
\bibinfo{author}{\bibfnamefont{T.}~\bibnamefont{deForest}} \bibnamefont{and}
  \bibinfo{author}{\bibfnamefont{J.~D.} \bibnamefont{Walecka}},
  \bibinfo{journal}{Adv. Phys.} \textbf{\bibinfo{volume}{15}},
  \bibinfo{pages}{1} (\bibinfo{year}{1966}).

\bibitem[{\citenamefont{Friar and Fallieros}(1984)}]{Fr84}
\bibinfo{author}{\bibfnamefont{J.~L.} \bibnamefont{Friar}} \bibnamefont{and}
  \bibinfo{author}{\bibfnamefont{S.}~\bibnamefont{Fallieros}},
  \bibinfo{journal}{Phys. Rev. C} \textbf{\bibinfo{volume}{29}},
  \bibinfo{pages}{1645} (\bibinfo{year}{1984}).

\bibitem[{\citenamefont{Machleidt et~al.}(1987)\citenamefont{Machleidt,
  Holinde, and Elster}}]{Ma87}
\bibinfo{author}{\bibfnamefont{R.}~\bibnamefont{Machleidt}},
  \bibinfo{author}{\bibfnamefont{K.}~\bibnamefont{Holinde}}, \bibnamefont{and}
  \bibinfo{author}{\bibfnamefont{C.}~\bibnamefont{Elster}},
  \bibinfo{journal}{Phys. Rep.} \textbf{\bibinfo{volume}{149}},
  \bibinfo{pages}{1} (\bibinfo{year}{1987}).

\bibitem[{\citenamefont{Arellano et~al.}(1996)\citenamefont{Arellano, Brieva,
  Sander, and von Geramb}}]{Ar96}
\bibinfo{author}{\bibfnamefont{H.~F.} \bibnamefont{Arellano}},
  \bibinfo{author}{\bibfnamefont{F.~A.} \bibnamefont{Brieva}},
  \bibinfo{author}{\bibfnamefont{M.}~\bibnamefont{Sander}}, \bibnamefont{and}
  \bibinfo{author}{\bibfnamefont{H.~V.} \bibnamefont{von Geramb}},
  \bibinfo{journal}{Phys. Rev. C} \textbf{\bibinfo{volume}{54}},
  \bibinfo{pages}{2570} (\bibinfo{year}{1996}).

\bibitem[{\citenamefont{Jacob and Wick}(1959)}]{Ja59}
\bibinfo{author}{\bibfnamefont{M.}~\bibnamefont{Jacob}} \bibnamefont{and}
  \bibinfo{author}{\bibfnamefont{G.~C.} \bibnamefont{Wick}},
  \bibinfo{journal}{Ann. Phys. (N.Y.)} \textbf{\bibinfo{volume}{7}},
  \bibinfo{pages}{404} (\bibinfo{year}{1959}).

\bibitem[{\citenamefont{Raynal}(1972)}]{Ra72}
\bibinfo{author}{\bibfnamefont{J.}~\bibnamefont{Raynal}}
  (\bibinfo{year}{1972}), \bibinfo{note}{saclay Note CEA-N-1529}.

\bibitem[{\citenamefont{Liu et~al.}(1996)\citenamefont{Liu, Stephenson, Bacher,
  Bowyer, Chang, Olmer, Wells, Wissink, and Lissantti}}]{Li96}
\bibinfo{author}{\bibfnamefont{J.}~\bibnamefont{Liu}},
  \bibinfo{author}{\bibfnamefont{E.~J.} \bibnamefont{Stephenson}},
  \bibinfo{author}{\bibfnamefont{A.~D.} \bibnamefont{Bacher}},
  \bibinfo{author}{\bibfnamefont{S.~M.} \bibnamefont{Bowyer}},
  \bibinfo{author}{\bibfnamefont{S.}~\bibnamefont{Chang}},
  \bibinfo{author}{\bibfnamefont{C.}~\bibnamefont{Olmer}},
  \bibinfo{author}{\bibfnamefont{S.~P.} \bibnamefont{Wells}},
  \bibinfo{author}{\bibfnamefont{S.~W.} \bibnamefont{Wissink}},
  \bibnamefont{and}
  \bibinfo{author}{\bibfnamefont{J.}~\bibnamefont{Lissantti}},
  \bibinfo{journal}{Phys. Rev. C} \textbf{\bibinfo{volume}{53}},
  \bibinfo{pages}{1711} (\bibinfo{year}{1996}).

\bibitem[{\citenamefont{Hicks et~al.}(1988)}]{Hi88}
\bibinfo{author}{\bibfnamefont{R.~S.} \bibnamefont{Hicks}}
  \bibnamefont{et~al.}, \bibinfo{journal}{Phys. Rev. Lett.}
  \textbf{\bibinfo{volume}{60}}, \bibinfo{pages}{905} (\bibinfo{year}{1988}).

\bibitem[{\citenamefont{Amos et~al.}(1989)\citenamefont{Amos, Koetsier, and
  Kurath}}]{Am89}
\bibinfo{author}{\bibfnamefont{K.}~\bibnamefont{Amos}},
  \bibinfo{author}{\bibfnamefont{D.}~\bibnamefont{Koetsier}}, \bibnamefont{and}
  \bibinfo{author}{\bibfnamefont{D.}~\bibnamefont{Kurath}},
  \bibinfo{journal}{Phys. Rev. C} \textbf{\bibinfo{volume}{40}},
  \bibinfo{pages}{374} (\bibinfo{year}{1989}).

\bibitem[{\citenamefont{Baghaei et~al.}(1992)}]{Ba92}
\bibinfo{author}{\bibfnamefont{H.}~\bibnamefont{Baghaei}} \bibnamefont{et~al.},
  \bibinfo{journal}{Phys. Rev. Lett.} \textbf{\bibinfo{volume}{69}},
  \bibinfo{pages}{2054} (\bibinfo{year}{1992}).

\end{thebibliography}

\end{document}